\documentclass{article}
\usepackage{kenmax,xspace,graphicx}
\graphicspath{{Eps/}}

\def\bdoc{\begin{document}}
\def\edoc{\end{document}}

\def\babs{\begin{abstract}}
\def\eabs{\end{abstract}}

\def\bbib{\begin{thebibliography}{999}}
\def\ebib{\end{thebibliography}}

\def\beq{\begin{equation}}
\def\eeq{\end{equation}}

\def\beqn{\begin{eqnarray}}
\def\eeqn{\end{eqnarray}}

\def\beqnn{\begin{eqnarray*}}
\def\eeqnn{\end{eqnarray*}}

\def\barr{\begin{array}}
\def\earr{\end{array}}

\def\bqu{\begin{quote}}
\def\equ{\end{quote}}

\def\bqun{\begin{quotation}}
\def\equn{\end{quotation}}

\def\bit{\begin{itemize}}
\def\eit{\end{itemize}}

\def\ben{\begin{enumerate}}
\def\een{\end{enumerate}}

\def\bpic{\begin{picture}}
\def\epic{\end{picture}}

\def\bfig{\protect \begin{figure}}
\def\efig{\protect \end{figure}}

\def\bcc{\begin{center}}
\def\ecc{\end{center}}

\def\brr{\begin{flushright}}
\def\err{\end{flushright}}

\def\bll{\begin{flushleft}}
\def\ell{\end{flushleft}}

\def\btab{\begin{tabular}}
\def\etab{\end{tabular}}

\def\ul{\underline}
\def\ol{\overline}

\def\bm#1{\mbox{\boldmath $#1$}}

\def\zerob{{\bm 0}}
\def\oneb{{\bm 1}}

\def\ab{{\bm a}}
\def\bb{{\bm b}}
\def\cb{{\bm c}}
\def\db{{\bm d}}
\def\eb{{\bm e}}
\def\fb{{\bm f}}
\def\gb{{\bm g}}
\def\hb{{\bm h}}
\def\ib{{\bm i}}
\def\jb{{\bm j}}
\def\kb{{\bm k}}
\def\lb{{\bm l}}
\def\mb{{\bm m}}
\def\nb{{\bm n}}
\def\ob{{\bm o}}
\def\pb{{\bm p}}
\def\qb{{\bm q}}
\def\rb{{\bm r}}
\def\sb{{\bm s}}
\def\tb{{\bm t}}
\def\ub{{\bm u}}
\def\vb{{\bm v}}
\def\wb{{\bm w}}
\def\xb{{\bm x}}
\def\yb{{\bm y}}
\def\zb{{\bm z}}

\def\Ab{{\bm A}}
\def\Bb{{\bm B}}
\def\Cb{{\bm C}}
\def\Db{{\bm D}}
\def\Eb{{\bm E}}
\def\Fb{{\bm F}}
\def\Gb{{\bm G}}
\def\Hb{{\bm H}}
\def\Ib{{\bm I}}
\def\Jb{{\bm J}}
\def\Kb{{\bm K}}
\def\Lb{{\bm L}}
\def\Mb{{\bm M}}
\def\Nb{{\bm N}}
\def\Ob{{\bm O}}
\def\Pb{{\bm P}}
\def\Qb{{\bm Q}}
\def\Rb{{\bm R}}
\def\Sb{{\bm S}}
\def\Tb{{\bm T}}
\def\Ub{{\bm U}}
\def\Vb{{\bm V}}
\def\Wb{{\bm W}}
\def\Xb{{\bm X}}
\def\Yb{{\bm Y}}
\def\Zb{{\bm Z}}

\def\alphab{\bm{\alpha}}
\def\betab{\bm{\beta}}
\def\deltab{\bm{\delta}}
\def\epsilonb{\bm{\epsilon}}
\def\gammab{\bm{\gamma}}
\def\omegab{\bm{\omega}}
\def\thetab{\bm{\theta}}
\def\xib{\bm{\xi}}
\def\lambdab{\bm{\lambda}}
\def\taub{\bm{\tau}}
\def\phib{\bm{\phi}}
\def\mub{\bm{\mu}}
\def\psib{\bm{\psi}}
\def\chib{\bm{\chi}}
\def\sigmab{\bm{\sigma}}

\def\Deltab{\bm{\Delta}}
\def\Lambdab{\bm{\Lambda}}
\def\Phib{\bm{\Phi}}
\def\Psib{\bm{\Psi}}
\def\Sigmab{\bm{\Sigma}}

\def\Ac{{\cal A}}
\def\Bc{{\cal B}}
\def\Cc{{\cal C}}
\def\Dc{{\cal D}}
\def\Ec{{\cal E}}
\def\Fc{{\cal F}}
\def\Gc{{\cal G}}
\def\Hc{{\cal H}}
\def\Ic{{\cal I}}
\def\Jc{{\cal J}}
\def\Kc{{\cal K}}
\def\Lc{{\cal L}}
\def\Mc{{\cal M}}
\def\Nc{{\cal N}}
\def\Oc{{\cal O}}
\def\Pc{{\cal P}}
\def\Qc{{\cal Q}}
\def\Rc{{\cal R}}
\def\Sc{{\cal S}}
\def\Tc{{\cal T}}
\def\Uc{{\cal U}}
\def\Vc{{\cal V}}
\def\Wc{{\cal W}}
\def\Xc{{\cal X}}
\def\Yc{{\cal Y}}
\def\Zc{{\cal Z}}

\def\wt#1{\widetilde{#1}}
\def\wh#1{\widehat{#1}}
\def\xh{\widehat{x}}
\def\thetah{\widehat{\theta}}
\def\betah{\widehat{\beta}}

\def\xbh{\widehat{\xb}}
\def\thetabh{\widehat{\thetab}}
\def\betabh{\widehat{\betab}}

\def\xbhk{\widehat{\xb}^{k}}
\def\thetahk{\widehat{\theta}^{k}}
\def\betahk{\widehat{\beta}^{k}}

\def\xbhkp{\widehat{\xb}^{k+1}}
\def\thetahkp{\widehat{\theta}^{k+1}}
\def\betahkp{\widehat{\beta}^{k+1}}

\def\thetabhk{\widehat{\thetab}^{k}}
\def\betabhk{\widehat{\betab}^{k}}

\def\thetabhkp{\widehat{\thetab}^{k+1}}
\def\betabhkp{\widehat{\betab}^{k+1}}

\def\thetamin{\theta_{\mbox{\tiny min}}}
\def\thetamax{\theta_{\mbox{\tiny max}}}
\def\betamin{\beta_{\mbox{\tiny min}}}
\def\betamax{\beta_{\mbox{\tiny max}}}

\def\ra{\rightarrow}
\def\la{\leftarrow}
\def\da{\downarrow}
\def\ua{\uparrow}

\def\Ra{\Rightarrow}
\def\La{\Leftarrow}
\def\Da{\Downarrow}
\def\Ua{\Uparrow}

\def\lra{\longrightarrow}
\def\lla{\longleftarrow}
\def\Lra{\Longrightarrow}
\def\Lla{\longleftarrow}

\def\lrarr{\leftrightarrow}
\def\Lrarr{\Leftrightarrow}
\def\udarr{\updownarrow}
\def\Uparr{\Updownarrow}

\def\d#1{\,\mbox{d}#1}
\def\dxdy{\d{x}\d{y}}
\def\dwxdwy{\d{\omega_x}\d{\omega_y}}
\def\dxdydz{\d{x}\d{y}\d{z}}

\def\disp#1{{\displaystyle #1}}
\def\diag#1{\mbox{diag}\left\{#1\right\}}

\def\Prob#1{\mbox{Pr}\left\{#1\right\}}
\def\var#1{\mbox{Var}\left\{#1\right\}}
\def\cov#1{\mbox{Cov}\left\{#1\right\}}
\def\corr#1{\mbox{Corr}\left\{#1\right\}}
\def\trace#1{\mbox{Tr}\left\{#1\right\}}
\def\rang#1{\mbox{rang}\left\{#1\right\}}
\def\det#1{\mbox{d\'et}\left\{#1\right\}}

\def\cosf{\cos \phi}
\def\sinf{\sin \phi}
\def\cost{\cos \theta}
\def\sint{\sin \theta}

\def\sgn{\mbox{sgn}}
\def\sinc{\mbox{sinc}}
\def\rect{\mbox{rect}}
\def\sincf#1{\mbox{sinc}\left(#1\right)}
\def\rectf#1{\mbox{rect}\left(#1\right)}
\def\trif#1{\mbox{tri}\left(#1\right)}
\def\xvec#1#2#3{\left\{#1_#2,\ldots,#1_#3\right\}}

\def\vx{\left[x_1,\ldots, x_n\right]^t}
\def\vz{\left[z_1,\ldots, z_n\right]^t}
\def\vw{\left[\omega_1,\ldots, \omega_n\right]^t}
\def\vxi{\left[\xi_1,\ldots, \xi_n\right]^t}
\def\iii{\int_{-\infty}^{+\infty}}
\def\izi{\int_{0}^{\infty}}
\def\izpi{\int_{0}^{\pi}}
\def\izdpi{\int_{0}^{2\pi}}
\def\intd{\int\kern-.8em\int}
\def\intt{\int\kern-.8em\int\kern-.8em\int}
\def\intg{\int\kern-1.1em\int}
\def\sumd{\mathop{\sum\sum}}

\def\sumi{\sum_{i=1}^{M}}
\def\sumj{\sum_{i=1}^{N}}
\def\sumk{\sum_{k=1}^{K}}
\def\sumn{\sum_{n=1}^{N}}
\def\summ{\sum_{m=1}^{M}}

\def\TA#1{{\cal A}\left\{ {#1} \right\}}
\def\TH#1{{\cal H}\left\{ {#1} \right\}}
\def\TP#1{{\cal P}\left\{ {#1} \right\}}
\def\TR#1{{\cal R}\left\{ {#1} \right\}}
\def\TRa#1{{\cal R}^{\dag}\left\{ {#1} \right\}}
\def\BR#1{{\cal B}\left\{ {#1} \right\}}
\def\TF#1{{\cal F}\left\{ {#1} \right\}}
\def\TFI#1{{\cal F}^{-1}\left\{ {#1} \right\}}
\def\TFn#1#2{{\cal F}_{#1}\left\{ {#2} \right\}}
\def\TFnI#1#2{{\cal F}_{#1}^{-1}\left\{ {#2} \right\}}
\def\Im#1{{\cal I}\mbox{m}\left(#1\right)}
\def\Ker#1{{\cal K}\mbox{er}\left(#1\right)}
\def\Imag#1{\mbox{Im}\left(#1\right)}
\def\Re#1{\mbox{Re}\left(#1\right)}
\def\expf#1{\exp\left[ {#1} \right]}

\def\dfdx#1#2{{\mbox{d} {#1}\over{\mbox{d} {#2}}}}
\def\dfdxd#1#2{{\mbox{d}^2 {#1}\over{\mbox{d} {#2}^2}}}
\def\dfdxt#1#2{{\mbox{d}^3 {#1}\over{\mbox{d} {#2}^3}}}
\def\dfdxn#1#2{{\mbox{d}^n {#1}\over{\mbox{d} {#2}^n}}}
\def\dfdxk#1#2{{\mbox{d}^k {#1}\over{\mbox{d} {#2}^k}}}

\def\dpdx#1#2{{{\partial {#1}\over \partial {#2}}}}
\def\dpdxd#1#2{{{\partial^2 {#1}}\over{\partial {#2}^2}}}
\def\dpdxdy#1#2#3{{{\partial ^2 {#1}}\over{\partial {#2} \partial {#3}}}}

\def\arg{\mbox{arg}}
\def\argmins#1#2{\mbox{arg}\min_{#1}\left\{{#2}\right\}}
\def\argmaxs#1#2{\mbox{arg}\max_{#1}\left\{{#2}\right\}}
\def\argmin#1#2{\mathop{\mbox{arg}\min}_{#1}\left\{{#2}\right\}}
\def\argmax#1#2{\mathop{\mbox{arg}\max}_{#1}\left\{{#2}\right\}}

\def\esp#1{\mbox{E}\left\{ #1 \right\}}
\def\espx#1#2{\mbox{E}_{#1}\left\{ #2 \right\}}

\def\wth#1{\widehat{\widetilde{\phantom{#1}}}\!\!\!\! #1}

\def\lrf{L_{r,\phi}}
\def\fw{\widehat{f}(\omegab)}
\def\fthwxi{\wth{f}(\Omega,\xib)}
\def\fthwfi{\wth{f}(\Omega,\phi)}
\def\ftrfi{\widetilde{f}(r,\phi)}

\def\fwxwy{\widehat{f}(\omega_x, \omega_y)}
\def\wxpwy{(\omega_x \, x + \omega_y \, y)}

\def\wtx{\omegab^t \cdot \xb}
\def\ejwtx{\exp\left[j \omegab^t \cdot \xb\right]}
\def\xitx{\xib^t \cdot \xb}

\def\ftrxi{\widetilde{f}(r,\xib)}
\def\ent{-\int p(x) \, \ln p(x) \d{x}}

\def\mean#1{\left< #1 \right>}
\def\slnhn{\sum_{n=1}^N \lambda_n h_n(\rb)}
\def\slngn{\sum_{n=1}^N \lambda_n g_n(\rb)}
\def\smngn{\sum_{n=1}^N \mu_n g_n(\rb)}
\def\slmhm{\sum_{m=1}^N \lambda_m h_m(\rb)}
\def\vlambda{\bm{\lambda} = [\lambda_1,\ldots,\lambda_n]}

\def\apriori{{\em a priori} }
\def\aposteriori{{\em a posteriori} }

\def\titre#1{\bcc{\Large\bf #1}\ecc}

\def\AMD{Ali Mohammad--Djafari}
\def\LSSa{Laboratoire des Signaux et Syst\`emes 
(CNRS--ESE--UPS) \\ 
\'Ecole Sup\'erieure d'\'Electricit\'e \\ 
Plateau de Moulon, 91192 Gif sur Yvette Cedex, France.}

\def\ME{maximum entropy}
\def\pdf{probability distribution function}
\def\lm{Lagrange multipliers}
\def\fix#1{\phi _#1(x)}
\def\fin{\fix n}
\def\fik{\fix k}
\def\fiz{\fix 0}
\def\sfinz{\sum_{n=0}^N \lambda_n \, \fin}
\def\sfinu{\sum_{n=1}^N \lambda_n \, \fin}
\def\bl{\bm{\lambda}}
\def\bd{\bm{\delta}}
\def\blz{\bl ^0}
\def\gnl{G _n(\bl)}
\def\gnlz{G _n(\blz)}
\def\un{n=1,\dots, N}
\def\nn{n=0,\dots, N}

\def\finn{\fin , \nn}
\def\esfinz{\exp\,\left[ -\sfinz \right] }
\def\esfinu{\exp\,\left[ -\sfinu \right] }
\def\esxm{\exp\,\left[ -\sum_{m=0}^N \lambda_m \, x^m \right] }
\def\efin{\esp \fin }
\def\zl{Z(\bl)}
\def\finxi{\phi _n(x_i)}
\def\snfinxi{\sum_{n=1}^N \lambda_n \finxi}
\def\esnfinxi{\exp \left[ - \snfinxi \right]}
\def\smfinxi{\sum_{i=1}^M \finxi}

\def\ejnw{\exp \left( -j n \omega_0 x \right) }
\def\eejnw{\mbox{E} \left\lbrace \ejnw \right\rbrace}

\def\signed#1{{\unskip\nobreak\hfil\penalty50\hskip2em\mbox{}
\nobreak\hfil\tt#1\parfillskip=0pt \finalhyphendemerits=0 \par}}

\def\uncatcodespecials{\def\do##1{\catcode`##1=12 }\dospecials}
\def\listing#1{\par\begingroup\setupverbatim\input#1 \endgroup}
\newcount\lineno
\def\setupverbatim{\tt \lineno=0
 \obeylines \uncatcodespecials \obeyspaces
 \everypar{\advance\lineno by1 \llap{\sevenrm\the\lineno\ \ }}}
{\obeyspaces\global\let =\ }

\def\defined{\stackrel{\mbox{def}}{=}}
\def\str{\stackrel}

\def\ER{\mbox{I\kern-.25em R}}
\def\EC{\mbox{C\kern-.8em C}}
\def\EZ{\mbox{Z\kern-.55em Z}}
\def\EN{\mbox{N\kern-.8em N}}

\def\singles{
 \abovedisplayskip 12pt plus 3pt minus 9pt
 \belowdisplayskip 12pt plus 3pt minus 9pt
 \abovedisplayshortskip 0pt plus 3pt
 \belowdisplayshortskip 7pt plus 3pt minus 4pt
 \baselineskip 14.4pt
 \lineskip 1pt
 \lineskiplimit 0pt}
\def\oneandhalf{
 \abovedisplayskip 18pt plus 3pt minus 9pt
 \belowdisplayskip 18pt plus 3pt minus 9pt
 \abovedisplayshortskip 0pt plus 3pt
 \belowdisplayshortskip 9.333pt plus 3pt
 \baselineskip 20pt
 \lineskip 2pt
 \lineskiplimit 1pt}

\def\double{
 \abovedisplayskip 24pt plus 3pt minus 9pt
 \belowdisplayskip 24pt plus 3pt minus 9pt
 \abovedisplayshortskip 0pt plus 3pt
 \belowdisplayshortskip 12pt plus 3pt
 \baselineskip 27pt
 \lineskip 3pt
 \lineskiplimit 2pt}

\def\dadb{\d{\alpha}\d{\beta}}

\def\ffbox#1{\fbox{\mbox{\vbox{#1}}}}

\def\rot{\mbox{rot}}
\def\case#1#2#3#4{
    \left\{
           \begin{array}{ll}
            {\displaystyle #1} & {\displaystyle #2} \cr 
            {\displaystyle #3} & {\displaystyle #4}
           \end{array}
    \right. }

\def\beqnarr#1&#2&#3\\#4&#5&#6\eeqnarr{
    \left\{
           \begin{array}{lcl}
            {\displaystyle #1} & #2 & {\displaystyle #3} \\ 
            {\displaystyle #4} & #5 & {\displaystyle #6} 
           \end{array}
    \right. }

\def\pyx{p(\yb|\xb)}
\def\pxy{p(\xb|\yb)}

\def\ie{{\em i.e.}}
\def\unsdpi{\left(\frac{1}{2\pi}\right)}
\def\unspi{\left(\frac{1}{\pi}\right)}
\def\up{\uppercase}
\def\zjm{z_{j-1}}
\def\zjp{z_{j+1}}
\def\fxyp{f(x,y)=\left\{
\barr{ll} 1 & (x,y)\in P\\ 0 & (x,y)\not\in P\earr
\right.}

\title{A Bayesian Approach for the Determination of the Charge
Density from Elastic Electron Scattering Data}
\author{
 A. Mohammad-Djafari\\
 Laboratoire des Signaux et Syst\`emes (CNRS-ESE-UPS),\\
 Plateau de Moulon, 91192 Gif-sur-Yvette, France\\
 and\\
 H. G. Miller\\
 Department of Physics, University of Pretoria,\\
 Pretoria 0002, South Africa
}

\date{\today}

\begin{document}
\maketitle

\index{Electron scattering}
\index{Inverse problems}
\index{Bayesian approach}
\index{Charge density}
\index{Fourier-Bessel}

\begin{abstract}
The problem of the determination of the charge density
from limited information about the charge form factor is
an ill-posed inverse problem.
A Bayesian probabilistic approach to this
problem which permits to take into account both errors and
prior information about the solution
is presented. We will show that many classical methods
can be considered as special cases of the proposed approach.
We address also the problem of the basis function
choice for the discretization and the uncertainty of the solution.
Some numerical results  for an analytical model
are presented to show the performance of
the proposed method.
\end{abstract}

\section{Introduction}

Elastic electron scattering provides a mean of determining the
charge density of a nucleus, $\rho(r)$, from the experimentally
determined charge form factor, $F(q)$.
The connection between the charge density and
the cross section is well understood and in plane wave Born
approximation $F(q)$ is just the Fourier transform of $\rho(r)$
which for the case of even-even nuclei, which we shall consider,
is simply given by
\beq\label{Fc}
F(q) = 4\pi\int_0^{\infty }r^2\,J_0(q r)\rho(r)\d{r}
\eeq
where $J_0$ is the spherical Bessel function of zero order and $q$
is the absolute value of the three momentum transfer.  Given that
the experimental measurements are performed over a limited range at
a finite number of values of the momentum transfer $q$, a unique
determination of $\rho(r)$ is not possible since the resulting
inverse problem is ill posed.

One of the generally accepted procedures for determining $\rho(r)$
is to expand it in a basis and then determine the expansion
coefficients from a least squares (LS) fit to the experimentally
measured values of $F(q)$
\cite{FN73,DFMRL74,HB83,Buck89,Macaulay94}.
The following questions then arise: how to
choose a basis and how to determine the order of the expansion?
Another problem with the LS methods is that increasing the number
of terms in the expansion generally leads to non physical oscillations
in the charge density in spite of the fact that the charge form
factor is well reproduced at the experimentally determined values
of $q$ \cite{W91,Miller96}.
Finally, due to the fact that the problem is inherently ill posed,
a small error in the data (experimental errors or measurement noise)
will produce large variations in the solution which is not acceptable
in practical situations.

What we are going to do is to show how a Bayesian approach can be
helpful to give both correct and reasonable answers to the
aforementioned questions and to propose new methods
which are more stable
with respect to the errors and finally to give procedures to put
the correct error bars on the proposed solutions.

\section{Fundamentals of the Bayesian approach}

\index{Minimum norm}
\index{Least squares}
\index{Maximum a posteriori}
\index{Maximum likelihood}

Let us start by discretizing the problem in the  usual
manner by expanding $\rho(r)$
in a basis $\phi_{n}(r)$:
\beq\label{Exp0}
\rho(r) =\left\{
\begin{array}{ll}
\sum_{n=1}^{N} a_n\phi_n(r) & r\leq R_{c}\\
0                           & r > R_{c}
\end{array}
         \right.
\eeq
and substituting it in (\ref{Fc}) yields
\beqn\label{Fcexp0}
F(q)
&=~& 4\pi\int_0^{R_{c}} r^2\,J_0(q r)\sum_{n=1}^{N} a_n\phi_n(r)\d{r}
\nonumber\\
&=~& 4\pi\sum_{n=1}^{N} a_n\int_0^{R_{c}} r^2\,J_0(q r)\phi_n(r)\d{r}
\eeqn
Now, defining
\beq\label{Amn}
A_{m,n}=4\pi\int_0^{R_{c}} r^2\,J_0(q_{m} r)\phi_n(r)\d{r}
\eeq
we obtain
\beq\label{Discrete0}
\bm{F}_{c}=\bm{A}\bm{a} +\bm{\epsilon}
\eeq
where $\bm{a}$ is a vector containing the coefficients
$\{a_{n}, n=1,\cdots,N\}$, $\bm{F}_{c}$ is a vector containing the
form factor data $\{F_{c}(q_{m}), m=1,\cdots,M\}$
and $A$ an $(M\times N)$ matrix containing the coefficients
$A_{m,n}$ given by (\ref{Amn}).
Note also that when the vector $\bm{a}$ is determined, we can calculate
$\bm{\rho}=\{\rho(r_k), \, k=1,\cdots,K\}$ by
\beq
\bm{\rho}=\bm{\Phi}\bm{a}
\eeq
where $\bm{\Phi}$ is a $K\times N$ matrix with the elements
$\Phi_{kn}=\phi_n(r_k)$.

The vector $\bm{\epsilon}$ is added to take account of the errors
in both measurement noise and due to discretization.
We assume that the components $\{\epsilon_{m}, m=1,\cdots,M\}$
are additive,
zero mean (no systematic error), mutually independent
(no correlation) and independent of $\bm{a}$,
and they can only be characterized by their
common variance $\sigma_{\epsilon}^{2}$.
This hypothesis is reasonable unless we  know more about its
characteristics.

Note that we have not yet discussed the choice of the basis
functions $\phi_{n}$ and  the determination of the expansion
order, $N$. We will come back to these questions later.
Let us  now see how  the Bayesian estimation approach works.

The main idea behind the Bayesian probabilistic approach is to
represent the uncertainty or any lack of knowledge or any diffuse
prior knowledge about a quantity by a probability law.
For example, the knowledge (or the hypothesis) that
$\{\bm{\epsilon}_{m}, m=1,\cdots,M\}$ are zero mean,
mutually independent and that they are only characterized by
their
common variance $\sigma_{\epsilon}^{2}$
can be described by choosing a Gaussian probability
distribution for them. One may also use the Maximum Entropy
Principle to enforce this choice. This means that we can write
\beq
p(\epsilon_m)=\frac{1}{(2\pi\sigma^{2})^{1/2}}
\exp\left\{-\frac{1}{2\sigma_{\epsilon}^{2}}\epsilon_{m}^{2}\right\}
\eeq
or
$p(\epsilon_{m})={\cal N}(0,\sigma_{\epsilon}^{2})$
and
\beq
p(\bm{\epsilon})=\prod_{m=1}^{M} p(\epsilon_{m})
=\frac{1}{(2\pi\sigma^{2})^{M/2}}
\exp\left\{-\frac{1}{2\sigma_{\epsilon}^{2}}\|\bm{\epsilon}\|^{2}\right\}
\eeq
or simply
$p(\bm{\epsilon}) = {\cal N}(\bm{0},\sigma_{\epsilon}^{2}\bm{I})$
where
$\bm{I}$ is the $(M\times M)$ unitary matrix and
$\sigma_{\epsilon}^{2}$
is the common variance of $\epsilon_{m}$ for all $m$.
Now, using this model (\ref{Discrete0}), we can define the
conditional
probability law
\beq\label{Likelihood}
p(\bm{F}_{c}|\bm{a}) =\frac{1}{(2\pi\sigma_{\epsilon}^{2})^{M/2}}
\exp\left\{
-\frac{1}{2\sigma_{\epsilon}^{2}}\|\bm{F}_{c}-\bm{A}\bm{a}\|^{2}
   \right\}
\eeq
or
$p(\bm{F}_{c}|\bm{a})={\cal
N}(\bm{A}\bm{a},\sigma_{\epsilon}^{2}\bm{I})$.
It is usual to call $p(\bm{F}_{c}|\bm{a})$ or its logarithm,
considered as a function of $\bm{a}$, the {\em Likelihood}.

One can stop here and  define the solution of the problem
(\ref{Discrete0}) as the vector $\wh{\bm{a}}$, which maximizes
the likelihood (ML):
\beq\label{ML1}
\wh{\bm{a}} =\arg\max_{\bm{a}}\left\{p(\bm{F}_{c}|\bm{a})\right\}
\eeq
or equivalently
\beq\label{ML2}
\wh{\bm{a}}
=\arg\min_{\bm{a}}\left\{-\ln p(\bm{F}_{c}|\bm{a})\right\}
\eeq
which, in the case of a Gaussian distribution (\ref{Likelihood})
becomes
\beq\label{LS0}
\wh{\bm{a}}
=\arg\min_{\bm{a}}\left\{\|\bm{F}_{c}-\bm{A}\bm{a}\|^{2})\right\}
\eeq
and we find here the LS solutions given by:
\beq\label{LS1}
(\bm{A}^t\bm{A}) \, \wh{\bm{a}}=\bm{A}^t\bm{F}_{c}
\eeq

The main problem with these solutions is that, very often, the matrix
$\bm{A}^t\bm{A}$ is either singular or at least ill-conditioned,
so that the solutions are very sensitive to the errors in the data
or even on the round-off errors during the numerical calculation.

The Bayesian approach can do better. In fact, before
looking at  the
data, we may have some  prior knowledge about $\bm{a}$.
For example, we know that $\bm{a}\in\ER^N$ and that we might
prefer those vectors which have a minimal norm $\|\bm{a}\|$.
To translate this prior knowledge and this preference, we may
assign a Gaussian probability distribution to the vector $\bm{a}$:
\beq\label{Prior1}
p(\bm{a}) =\frac{1}{(2\pi\sigma_{a}^{2})^{M/2}}
\exp\left\{ -\frac{1}{2\sigma_{a}^{2}}\|\bm{a}\|^{2}\right\}
\eeq
where $\sigma_{a}^{2}$ gives an idea
about the scale of the norm of
the vector $\bm{a}$.
Now, using the Bayes rule we can calculate the posterior probability
law
\beq\label{Bayes}
p(\bm{a}|\bm{F}_{c})
=\frac{p(\bm{F}_{c}|\bm{a})\, p(\bm{a})}{p(\bm{F}_{c})}
\eeq
where the denominator
\beq\label{Evidence1}
p(\bm{F}_{c})=\intg p(\bm{F}_{c}|\bm{a})\, p(\bm{a})\d{\bm{a}}
\eeq
is the normalization constant (called sometimes the {\em Evidence})
\cite{Skilling89}.

This posterior law contains all the information we may wish about the
solution. For example, we may want to know what is the probability
that $\underline{\bm{a}}<\bm{a}\le\overline{\bm{a}}$.
This can be calculated by
\beq
P(\underline{\bm{a}}<\bm{a}\le\overline{\bm{a}})
=\intg_{\underline{\bm{a}}}^{\overline{\bm{a}}}
p(\bm{a}|\bm{F}_{c})\d{\bm{a}}
\eeq

Or, we may be interested only in one of these parameters
$a_{n}$ and want to know what is the probability
that $\underline{a}_{n}<a_{n}\le\overline{a}_{n}$.
This can be calculated by
\beq
P(\underline{a}_{n}<a_{n}\le\overline{a}_{n})
=\int_{\underline{a}_{n}}^{\overline{a}_{n}}
p(a_{n}|\bm{F}_{c})\d{a}_{n}
\eeq
where the marginal posterior law $p(a_{n}|\bm{F}_{c})$ can be
calculated by
\beq\label{MarginalPosterior}
p(a_{n}|\bm{F}_{c})=\int\cdots\int p(\bm{a}|\bm{F}_{c})
\d{a}_{1}\cdots\d{a}_{n-1}\d{a}_{n+1}\cdots\d{a}_{N}.
\eeq
We can also simply define as the solution the vector $\wh{\bm{a}}$
which corresponds to the mean value of the posterior
law~--called Posterior mean (PM) estimator:
\beq\label{PosteriorMean}
\wh{\bm{a}}
=\intg\bm{a}\, p(\bm{a} |\bm{F}_{c})\d{\bm{a}}
\eeq
or the vector $\wh{\bm{a}}$
which maximizes this posterior distribution~--called Maximum a
posteriori (MAP) estimator:
\beq\label{MAP}
\wh{\bm{a}}
=\arg\max_{\bm{a}}\left\{p(\bm{a} |\bm{F}_{c})\right\}
=\arg\min_{\bm{a}}\left\{-\ln p(\bm{a} |\bm{F}_{c})\right\}
\eeq
or even the vector $\wh{\bm{a}}$ whose components $\wh{a}_{n}$
correspond to the maximizer of the marginal posterior law
(\ref{MarginalPosterior})
--called Marginal MAP estimator:
\beq\label{MarginalMAP}
\wh{a}_{n}=\arg\max_{a_{n}} \left\{p(a_{n}|\bm{F}_{c})\right\}
=\arg\min_{a_{n}} \left\{-\ln p(a_{n}|\bm{F}_{c})\right\}.
\eeq

In the following we consider only the MAP estimator (\ref{MAP}).
Using the probability distributions (\ref{Likelihood}) and
(\ref{Prior1}) in (\ref{Bayes}), the MAP solution is given by:
\beq\label{MNLS0}
\wh{\bm{a}}=\arg\min_{\bm{a}}
\left\{\|\bm{F}_{c}-\bm{a}\|^{2}+\lambda\|\bm{a}\|^{2})\right\}
\eeq
where $\lambda=\left(\sigma_{\epsilon} /\sigma_{a}\right)^{2}$ and
we find the minimum norm least squares (MNLS) solution which is
given  explicitly by
\beq\label{MNLS1}
\wh{\bm{a}}
=(\bm{A}^t\bm{A}+\lambda\bm{I})^{-1}\bm{A}^t\bm{F}_{c}.
\eeq
Comparing (\ref{LS1}) and (\ref{MNLS1}) we see that, for a
given $N$, the matrix $(\bm{A}^t\bm{A}+\lambda\bm{I})$ is
always better conditioned than the matrix $(\bm{A}^t\bm{A})$
and so the solution (\ref{MNLS1}) always is more stable than
the solution (\ref{LS1}).

We may also want some information about the uncertainty of this solution.
For this we can use the posterior law $p(\bm{a}|\bm{F}_{c})$.
For example, using the likelihood (\ref{Likelihood}) and the prior law
(\ref{Prior1}), it is easy to show that the posterior law is
Gaussian, i.e.
$p(\bm{a}|\bm{F}_{c})={\cal N}(\wh{\bm{a}},\wh{\bm{P}})$
with $\wh{\bm{a}}$ given by (\ref{MNLS1}) and
$\wh{\bm{P}}=(\bm{A}^t\bm{A}+\lambda\bm{I})^{-1}$.
We can then use the diagonal elements of the posterior covariance
matrix $\wh{\bm{P}}$ to calculate the posterior variances of the
estimates, i.e. $\mbox{Var}(a_{n})=\wh{P}_{nn}$
and so put the error bars on the solution.
When the posterior law is not Gaussian, we can always
calculate
\beq\label{Mean}
\mbox{E}(a_{n})=\int a_{n} \, p(a_{n}|\bm{F}_{c}) \d{a}_{n}
\eeq
and
\beq\label{Var}
\mbox{Var}(a_{n})
=\int \left(a_{n}-\mbox{E}(a_{n})\right)^{2} \,
p(a_{n}|\bm{F}_{c}) \d{a}_{n}
\eeq
but in general we may not have explicit expressions for these
integrals. We can however do numerical calculation either
by approximating the posterior law by a Gauusian
law or by a stochastic integral calculation.

One question still remains: how to determine $\lambda$ and $N$?
Three approaches are possible:

\ben
\item Assign them experimentally from the data using some knowledge
on the physics of the problem. For example, the Parseval-type
relation between $\rho(r)$ and $F_{c}(q)$:
\beq\label{Parseval1}
\int 4\pi r^{2}\rho^{2}(r)\d{r}
=\frac{1}{(2\pi)^{3}}\int 4\pi q^{2} Fc(q)\d{q}
\eeq
can be used to estimate $\sigma_{a}^{2}$ by:
\beq\label{Parseval2}
\sigma_{a}^{2}
=\frac{1}{N}\sum_{n=1}^{N} a_{n}^{2}
=\frac{1}{M}\sum_{m=1}^{M} F_{c_{m}}^{2}
\eeq
and having an estimate of the noise variance $\sigma_{\epsilon}^{2}$
we can determine $\lambda$.

\item Consider $\lambda$ and $N$ as two extra parameters
(hyper parameters) to estimate jointly with the unknown
parameter $\bm{a}$. We can then assign a prior law
for them. For example Jeffrey's priors $p(\lambda)=\frac{1}{\lambda}$
for $\lambda$ and a uniform $p(N)=1/N_{\mbox{\tiny max}}$ for $N$.
(Other choices are possible, for example a Gamma prior $\lambda$
which eliminates $\lambda=0$ and $\lambda=\infty$ and
a binomial prior for $N$ which eliminates $N=0$ and
$N=N_{\mbox{\tiny max}}$.)

Finally, we can estimate them jointly with $\bm{a}$ by
\beq\label{GML}
(\wh{\bm{a}},\wh{\lambda},\wh{N})
=\arg\max_{(\bm{a},\lambda,N)}
\left\{p(\bm{a},\lambda,N|\bm{F}_{c})\right\}
\eeq
where
\beq
p(\bm{a},\lambda,N|\bm{F}_{c})\propto
p(\bm{F}_{c}|\bm{a},\lambda,N)\, p(\bm{a}|\lambda,N)
\, p(\lambda)\, p(N).
\eeq
We must however be careful to verify that this joint criterion has
at least a local optimum.

\item Consider $\lambda$ and $N$ as two extra parameters as in the
precedent case but not on the same level. This means that we can try
to estimate them first by
\beq\label{MML}
(\wh{\lambda},\wh{N})
=\arg\max_{(\lambda,N)}\left\{ p(\lambda,N|\bm{F}_{c})\right\}
\eeq
where
\beq\label{MML1}
p(\lambda,N|\bm{F}_{c})=\intg p(\bm{a},\lambda,N|\bm{F}_{c})\d{\bm{a}}
\eeq
and then use them in (\ref{MAP}).
Note, however that finding an analytical expression for
$p(\lambda,N|\bm{F}_{c})$ is not always possible and its numerical
calculation may be very costly.

\item Consider $\lambda$ and $N$ as two nuisance parameters,
integrate them out and estimate $\bm{a}$ directly by
\beq\label{MMAP}
\wh{\bm{a}}
=\arg\max_{\bm{a}}\left\{p(\bm{a}|\bm{F}_{c})\right\}
=\arg\max_{\bm{a}}\left\{
\sum_{n=1}^{N}\intg p(\bm{a},\lambda,n|\bm{F}_{c})\d{\lambda}
\right\}
\eeq
\een
(For more details on these methods, their relative characteristics,
their practical implementations and their relatives performances
see \cite{Djafari90,MacKay93,Djafari94,Djafari95}).

Still one question remains: the choice of the basis-functions.

\section{Choice of the basis-functions}

Two approaches are used to select the basis functions.
We call them the
{\em operator based parametric} approach and the
{\em non parametric} approach and
we will discuss both in detail in the following sections.
We propose then a new third approach which tries to eliminate the
limitations and to keep the advantages of the previous approaches.
We call this third approach {\em physically based parametric}.

\subsection{Operator based parametric approach}

The first approach  is to choose special purpose basis functions
based on the properties of the operator linking the data to the
unknowns. For example in our case, due to the fact that the kernel of
the integral operator of the direct problem is a Bessel function, we
may also use the Bessel functions as the basis functions for $\rho(r)$
\beq\label{Exp1}
\rho(r) =\left\{
\begin{array}{ll}
\sum_{n=1}^{N} a_n j_0(q_n r) & r\leq R_{c}\\
0                             & r > R_{c}
\end{array}
         \right.
\eeq
This will permit us, using the orthogonality relation
\beq\label{FBOrthogonality}
\int_{0}^{R_{c}} r^{2} \, J_{l}(q_{n} r) \, J_{l}(q_{m} r) \d{r}
=\frac{R_{c}^{3}}{2} \,  J_{l+1}^{2}(q_{n} R_{c}) \, \delta_{n,m},
\eeq
to find an explicit expression for the charge form factor as a
function of the coefficients $a_{n}$ :
\beq\label{Fcexp1}
F(q)=\frac{4\pi R_{c}^{2}}{q} \sum_{n=1}^{N} a_{n} \, \frac{(-1)^{n}}
{(q R_{c})^2 -(n\pi)^2} \, \sin(q R).
\eeq

With this choice, note also that,
if the form factor $F(q)$ was known exactly at
$q_{n}=\frac{n\pi}{R_{c}}$
then the coefficients $a_{n}$ could be
calculated analytically by
\beq\label{Coeff}
a_{n}=\frac{F(q_{n})}{2\pi R_{c}^{3} \left[J_{1}(q_{n} R)\right]^{2}}.
\eeq

In general, however, the cross section is measured at momentum
transfers different from $q_{n}=\frac{n\pi}{R_{c}}$.

Now, assume that we are given $M$ measurements at arbitrary
momentum transfers $\bm{q}=\{q_{1},q_{2}, \ldots, q_{M}\}$
and we wish to
determine the $N$ expansion coefficients
$\bm{a}=\{a_{1},a_{2},\ldots, a_{N}\}$.
In this case Eq.\ (\ref{Fcexp1}) leads to
\beq\label{Discrete1}
\bm{F}_{c} =\bm{A}\bm{a} +\bm{\epsilon}
\eeq
as in (\ref{Discrete0}).

The main advantage of this approach is the fact that $\bm{a}$ is
a small dimension vector and so is the matrix $\bm{A}$ and we have
an explicit analytical expression for calculating its elements.

But at least one main
disadvantage to  such a choice is that our prior knowledge on
$\bm{a}$ may be limited. For example, if we know that $\rho(r)$
is a positive function we cannot easily incorporate this
information in the parameters, $\bm{a}$.

\subsection{Non parametric approach}

The second approach is to choose the basis-functions as general as
possible and independently of the direct problem operator, for
example, either:
\beq\label{Basis0}
\phi_{n}(r)=\delta(r-n\Delta)
\eeq
or
\beq\label{Basis1}
\phi_{n}(r)=\left\{\barr{ll}
1 &\mbox{if~} (n-1)\Delta< r\le n\Delta\\
0 & {\mbox elsewhere}
\earr\right.
\eeq
with $\Delta$ chosen appropriately small (maximum needed resolution)
to be certain we are able to approximate any function $\rho(r)$ as
precisely as desired.
But, this means that $N$ will probably be large.
This is a disadvantage, but this can be compensated, as we will
see further, by the fact that  the coefficients $a_{n}$ now have a
direct physical meaning:
the samples of $\rho(r)$ in the first case and the mean
values of $\rho(r)$ in the intervals $(n-1)\Delta< r\le n\Delta$
in the second case. This means, for example, that the prior knowledge
such as the smoothness or the positivity of the function $\rho(r)$
can be transmitted to the coefficients $a_{n}$ easily.

Let us choose (\ref{Basis0}) and go further into the  details.
Replacing (\ref{Exp0}) with the basis-functions (\ref{Basis0}) in
(\ref{Fc}) we obtain:
\beq\label{Fcexp2}
F(q) =\sum_{n=1}^{N} a_{n}\int_0^{R_{c}}\d{r}\,4\pi r^2\,J_0(q r)
\delta (r-n\Delta)
=4\pi \sum_{n=1}^{N} a_{n} \Delta\,(n\Delta)^2\,J_0(n\Delta q)
\eeq
Denoting by
\beq\label{Amn1}
A_{m,n}=4\pi \, \Delta\,(n\Delta)^2\, J_0(n\Delta q_{m})
\eeq
we obtain $\bm{F}_{c}=\bm{A}\bm{a}+\bm{\epsilon}$ as in
(\ref{Discrete1}).
If we use (\ref{Basis1}) in place of (\ref{Basis0}), the only change
will be in the expression of the Matrix elements $A_{m,n}$ which
become
\beq\label{Amn2}
A_{m,n}=4\pi\int_{(n-1)\Delta}^{n\Delta}\,r^2\,J_0(q_{m} r)\d{r}.
\eeq

To make a distinction between this approach and the preceding one, let us
denote
$\bm{a}$ by $\bm{\rho}$ and $\bm{A}$ by $\bm{B}$:
\beq\label{Discrete2}
\bm{F}_{c} =\bm{B}\bm{\rho} +\bm{\epsilon}
\eeq

Let us now compare  (\ref{Discrete2}) and (\ref{Discrete1}):
$\bm{a}$ in~(\ref{Discrete1}) is a  vector of small dimension while
$\bm{\rho}$ in~(\ref{Discrete2}) is a vector of much larger
dimension.

Here  we can also define either the LS solution:
\beq\label{LS2}
\wh{\bm{\rho}}=\arg\min_{\bm{\rho}}
\left\{\|\bm{F}_{c}-\bm{B}\bm{\rho}\|^{2}\right\}
= (\bm{B}^t\bm{B})^{-1}\bm{B}^t\bm{F}_{c}
\eeq
or the MNLS solution:
\beq\label{MNLS2}
\wh{\bm{\rho}}=\arg\min_{\bm{\rho}}
\left\{\|\bm{F}_{c}-\bm{B}\bm{\rho}\|^{2}+\lambda\|\bm{\rho}\|^{2}
\right\}
= (\bm{B}^t\bm{B}+\lambda\bm{I})^{-1}\bm{B}^t\bm{F}_{c}
\eeq
but neither of these solutions may be satisfactory.

In~(\ref{Discrete2}) it is possible to incorporate the smoothness
and the positivity of the function $\rho(r)$ into a more appropriate
prior distribution  for the components $\rho_{n}$.
For example, to enforce the smoothness of $\rho(r)$ we can assign
\beq
p(\bm{\rho})=p(\rho_{1})\prod_{n=2}^{N} p(\rho_{n}|\rho_{n-1})
={\cal N}(\rho_{n-1},\sigma_{0}^2)
\eeq
with
\beq
p(\rho_{1})={\cal N}(\rho_{0},\sigma_{0}^2)
\propto
\exp\left[-\frac{1}{2\sigma_{0}^2}(\rho_{1}-\rho_{0})^{2}\right]
\eeq
\beq
p(\rho_{n}|\bm{\rho})=p(\rho_{n}|\rho_{n-1})
={\cal N}(\rho_{n-1},\sigma_{0}^2)
\propto
\exp\left[-\frac{1}{2\sigma_{0}^2}(\rho_{n}-\rho_{n-1})^{2}\right]
\eeq
which leads to
\beq\label{Prior2}
p(\bm{\rho})
\propto\exp\left[
-\frac{1}{2\sigma_{0}^2}\sum_{n=1}^{N}(\rho_{n}-\rho_{n-1})^{2}
\right]
\eeq
Using this prior distribution in (\ref{Bayes}), the MAP estimator
becomes
\beq\label{MAP2}
\wh{\bm{\rho}}=\arg\max_{\bm{\rho}}\{p(\bm{\rho}|\bm{F}_{c})\}
=\arg\min_{\bm{\rho}}\{J(\bm{\rho})\}
\eeq
with
\beq\label{Crit1}
J(\rho)=\|\bm{F}_{c}-\bm{B}\bm{\rho}\|^{2}
+\lambda\sum_{n=1}^{N} (\rho_{n}-\rho_{n-1})^{2}
\eeq
Defining the matrix
\beq
\bm{D}=\pmatrix{
1 & -1\cr
& 1 & -1\cr
& &\ddots &\ddots\cr \cr
& & & & 1 & -1}
\eeq
it is easy to show that
\beq\label{Crit2}
J(\rho)=\|\bm{F}_{c}-\bm{B}\bm{\rho}\|^{2}
+\lambda \, \|(\bm{D}\bm{\rho}\|^{2}+\lambda \, (\rho_{1}-\rho_{0})^{2}
\eeq
Let us temporarily assume that  $\rho_{1}=\rho_{0}$.
We then  have an explicit
solution for the minimizer of (\ref{Crit2}) which is given by:
\beq\label{MAP2explicite}
\wh{\bm{\rho}}
=(\bm{B}^t\bm{B}+\lambda\bm{D}^t\bm{D})^{-1}\bm{B}^t\bm{F}_{c}
\eeq
Comparing this solution with the MNLS solution (\ref{MNLS2})
gives us the possibility to see the difference in which  the term
$\bm{D}^t\bm{D}$ is used in place of $\bm{I}$. Indeed, due to the
fact that $\bm{D}$ corresponds to a first order derivative, we may
designate
the MNLS solution as the zero order regularized solution in contrast
to  the first order regularized solution.
It is possible to extend this to more general regularized
solutions by  an appropriate choice of the matrix $\bm{D}$.

Now, let us go back to (\ref{Crit2}). $\rho_{0}$ is now a
new extra hyper-parameter which may play a great role in the solution
of our inverse problem where the data do not contain
information about the DC  level of the function $\rho(r)$.

One way to enforce the positivity of the solution is to choose a
prior
distribution  such as :
\beq\label{Prior3}
p(\bm{\rho})=\sum_{n=1}^{N} p(\rho_{n})
=\sum_{n=1}^{N}\frac{\beta^{\alpha}}{\Gamma(\alpha)} \, \rho_{n}^{\alpha-1}
\exp[-\beta\rho_{n}]
\eeq
which can also be written as:
\beq
p(\bm{\rho})\propto\exp\left[
-\sum_{n=1}^{N} (1-\alpha)\ln\rho_{n}+\beta\rho_{n}
\right]
\eeq
and which is  called an {\em Entropic prior} in \cite{Djafari90}.
Using this prior law in (\ref{Bayes}), the MAP estimator becomes
\beq\label{MAP3}
\wh{\bm{\rho}}=\arg\max_{\bm{\rho}}\{p(\bm{\rho}|\bm{F}_{c})\}
=\arg\min_{\bm{\rho}}\{J(\bm{\rho})\}
\eeq
with
\beq\label{Crit3}
J(\bm{\rho})=
\|\bm{F}_{c}-\bm{B}\bm{\rho}\|^{2}
+\lambda_{1}\sum_{n=1}^{N}\ln\rho_{n}
+\lambda_{2}\sum_{n=1}^{N}\rho_{n}
\eeq
where $\lambda_{1}$ and $\lambda_{2}$ are related to $\alpha$,
$\beta$ and $\sigma_{\epsilon}^{2}$.
Other choices are possible \cite{Djafari90}.

To enforce both positivity and the smoothness we propose here to
choose
\beq\label{Prior4}
p(\bm{\rho})\propto\exp\left[
-\lambda_{1}\sum_{n=1}^{N} (\rho_{n}-\rho_{n-1})^{2}
-\lambda_{2}\sum_{n=1}^{N}\ln\rho_{n}
-\lambda_{3}\sum_{n=1}^{N}\rho_{n}
\right]
\eeq
which leads to
\beq\label{MAP4}
\wh{\bm{\rho}}=\arg\min_{\bm{\rho}>0}\left\{J(\bm{\rho})\right\}
\eeq
with
\beq\label{Crit4}
J(\bm{\rho})
=\|\bm{F}_{c}-\bm{B}\bm{\rho}\|^{2}
+\sum_{n=1}^{N}\lambda_{1} (\rho_{n}-\rho_{n-1})^{2}
+\lambda_{2}\ln\rho_{n} +\lambda_{3}\rho_{n}
\eeq

\subsection{Physically based parametric approach}

In this approach we choose special purpose basis functions
based on the physics of the problem. For example, in our case,
since the charge density is a single-valued function defined in a
finite domain,
the Fourier-Bessel (FB) basis functions which satisfy both the
orthogonality and the concentration property conditions, can be used
for expansion:

\index{Fourier-Bessel}

\beq\label{Exp2}
\rho(r) =\left\{
\begin{array}{ll}
\sum_{n=1}^{N} a_n j_0(q_n r) & r\leq R_{c}\\
0                             & r > R_{c}
\end{array}
         \right.
\eeq
where $q_{n} =\frac{n\pi}{R_{c}}$.
Excepted the original motivation, this choice is exactly the same
as in the first approach and all the relations developed and discussed
there can be used.

We may choose other basis functions which are more
appropriate to translate our prior knowledge on the desired
solution. For example, in our case, we know a priori that,
the solution is smooth, positive and a decreasing function.
Then we can choose the following function :
\beq\label{Exp3}
\rho(r) =\left\{
\begin{array}{ll}
\sum_{n=1}^{N} a_n \exp(-q_n r^2) & r\leq R_{c}\\
0                               & r > R_{c}
\end{array}
         \right.
\eeq
Using this expansion in eq.\ (\ref{Fc})
we find $\bm{F}_{c}=\bm{A}\bm{a}+\bm{\epsilon}$ as in
(\ref{Discrete0}) or in (\ref{Discrete1}), where
\beq\label{Amn3}
A_{m,n}=4\pi \int_0^{R_{c}}\d{r}\, r^2\,J_0(q_{m} r)\exp(-q_{n} r^2).
\eeq

With this choice we keep the main advantage of the first approach
which is the small dimension of the vector $\bm{a}$ and the main
advantages of the second approach which is the translation of our
prior knowledge of the positivity of the function $\rho(r)$.
This is due to the fact that if we impose the positivity constraint on
the coefficients $a_{n}$ we insure that the solution remains always
positive.

In the next section we will illustrate the performance of these
different solutions for the estimation of the charge density from
elastic electron scattering data.

\section{Issues on the uncertainty of the solution}

In any scientific problem solving, a proposed solution should be given
in any way with a measure of its uncertainty or confidence.
In Bayesian approach, the posterior probability gives us naturally the
necessary tool.
To see this, let come back to our problem and make a summary.
We have a set of data $\bm{F}$ and we want to estimate $\rho(r)$ or more
precisely $\bm{\rho}$ for some locations $r_i$.
Let assume that we have chosen a constant discretization step and so,
we want to estimate a vector $\bm{\rho}=\{\rho(r_i)\, i=1,\cdots,K\}$.

We presented two approaches: parametric and non-parametric.
In the first approach, we have
\beqn
\bm{F}   &=~& \bm{A} \bm{a} + \bm{\epsilon} \\
\bm{\rho} &=~& \bm{\Phi} \bm{a}
\eeqn
and in the second
\beqn
\bm{F}   = \bm{B} \bm{\rho} + \bm{\epsilon}
\eeqn
In both cases, we are interested to $\bm{\rho}$.
In the first approach, we assigned $p(\bm{a})$ and $p(\bm{F}|\bm{a})$,
calculated the posterior $p(\bm{a}|\bm{F})$, defined a solution
$\widehat{\bm{a}}$ for the parameters $\bm{a}$, and finally, a solution
$\widehat{\rho}=\bm{\Phi}\widehat{\bm{a}}$ for $\bm{\rho}$.
In the second, we assigned directly $p(\bm{\rho})$ and $p(\bm{F}|\bm{\rho})$,
calculated the posterior $p(\bm{\rho}|\bm{F})$, and finally defined a solution
$\widehat{\bm{\rho}}$.
In both cases, we can use the posterior laws to quantify the uncertainty
of the solutions.
There are, at least, three approaches:
\bit
\item Simply generate samples from the posterior law $p(\bm{\rho}|\bm{F})$
using for example a monte carlo method, and show all these samples to
see the distribution of the proposed solution.

\item Calculate the posterior mean and the posterior variance of the
solution at each
point either analytically (when possible) or numerically using for
example the samples generated by a monte carlo method.

\item Calculate the posterior mean and covariance of the
solution either analytically (when possible) or approximate it
numerically by any quadrature algorithm.
\eit

To illustrate this, let consider the cases where all the probability laws
are Gaussian. Then, all the calculations can be done analytically.
The following summarizes all the steps for the calculation of the
solutions and their posterior covariances in the above-mentioned two
cases:
{\small
\[
\barr{|c|c|} \hline\hline & ~\\
\hbox{Non-parametric} & \hbox{Parametric} \\ & ~\\ \hline
\barr{@{}l@{}l@{}l@{}}
&& \\ ~\\
\bm{F} &=~& \bm{B}\bm{\rho}+\bm{\epsilon}
\\ ~ \\  ~\\  \hline ~\\
p(\bm{\rho}) &=~& {\cal N}(\bm{\rho}_0,\sigma_{\rho}^2 \bm{P}_0) \\ ~\\
p(\bm{F}|\bm{\rho}) &=~& {\cal N}(\bm{B}\bm{\rho},\sigma_{\epsilon}^2 \bm{I})
\\ ~\\ \hline ~\\
p(\bm{\rho}|\bm{F}) &=~& {\cal N}(\wh{\bm{\rho}},\bm{P}_{\rho}) \\ ~\\
\wh{\bm{\rho}} &=~& [\bm{B}^t\bm{B}+\lambda\bm{P}_0^{-1}]^{-1}\bm{B}^t(\bm{F}-\bm{B}\bm{\rho}_0)
\\ ~\\
\bm{P}_{\rho} &=~& [\bm{B}^t\bm{B}+\lambda\bm{P}_0^{-1}]^{-1} \\ ~\\
&& \hbox{~with~} \lambda=\sigma_{\epsilon}^2/\sigma_{\rho}^2
\\ ~ \\ ~\\ ~\\ \hline ~\\
\multicolumn{3}{c}{\hbox{Special cases}}\\ ~\\
\sigma_{\epsilon}\longrightarrow 0 &&
\left\{\barr{@{}l@{}}
\wh{\bm{\rho}}= [\bm{B}^t\bm{B}]^{-1}\bm{B}^t \bm{F} \\ ~\\
\bm{P}_{\rho}=\bm{0}
\earr\right.
\\ ~\\
\sigma_{\rho}\longrightarrow 0 &&
\left\{\barr{@{}l@{}}
\wh{\bm{\rho}}= \bm{\rho}_0 \\ ~\\
\bm{P}_{\rho}=\bm{0}
\earr\right.
\\ ~\\
\sigma_{\epsilon}\longrightarrow \infty &&
\left\{\barr{@{}l@{}}
\wh{\bm{\rho}}= \bm{\rho}_0 \\ ~\\
\bm{P}_{\rho}=\sigma_{\rho}^2\bm{P}_0
\earr\right.
\\ ~\\
\sigma_{\rho}\longrightarrow \infty &&
\left\{\barr{@{}l@{}}
\wh{\bm{\rho}}= [\bm{B}^t\bm{B}]^{-1}\bm{B}^t \bm{F} \\ ~\\
\bm{P}_{\rho}=\sigma_{\epsilon}^2 [\bm{B}^t\bm{B}]^{-1}
\earr\right.
\\
\earr
&
\barr{@{}l@{}l@{}l@{}}
&& \\
\bm{F}   &=~& \bm{A}\bm{a}+\bm{\epsilon} \\ ~\\
\bm{\rho} &=~& \bm{\Phi}\bm{a}
\\ ~\\ \hline ~\\
p(\bm{a}) &=~& {\cal N}(\bm{a}_0,\sigma_{a}^2 \bm{I}) \\ ~\\
p(\bm{F}|\bm{a}) &=~& {\cal N}(\bm{A}\bm{a},\sigma_{\epsilon}^2 \bm{I})
\\ ~\\ \hline ~\\
p(\bm{a}|\bm{F}) &=~& {\cal N}(\wh{\bm{a}},\bm{P}_{a}) \\ ~\\
\wh{\bm{a}} &=~& [\bm{A}^t\bm{A}+\lambda\bm{I}]^{-1}\bm{A}^t(\bm{F}-\bm{A}\bm{a}_0) \\ ~\\
\bm{P}_{a} &=~& [\bm{A}^t\bm{A}+\lambda\bm{P}_0^{-1}]^{-1} \\ ~\\
&& \hbox{~with~} \lambda=\sigma_{\epsilon}^2/\sigma_{a}^2 \\ ~\\
p(\bm{\rho}|\bm{F}) &=~&
{\cal N}(\bm{\Phi}\wh{\bm{a}},\bm{\Phi}\bm{P}_{a}\bm{\Phi}^t)
\\ ~\\ \hline ~\\
\multicolumn{3}{c}{\hbox{Special cases}} \\ ~\\
\sigma_{\epsilon}\longrightarrow 0 &&
\left\{\barr{@{}l@{}}
\wh{\bm{\rho}}= \bm{\Phi} [\bm{A}^t\bm{A}]^{-1}\bm{A}^t \bm{F} \\ ~\\
\bm{P}_{\rho}=\bm{0}
\earr\right.
\\ ~\\
\sigma_{a}\longrightarrow 0 &&
\left\{\barr{@{}l@{}}
\wh{\bm{\rho}}= \bm{\Phi}\bm{a}_0 \\ ~\\
\bm{P}_{\rho}=\bm{0}
\earr\right.
\\ ~\\
\sigma_{\epsilon}\longrightarrow \infty &&
\left\{\barr{l}
\wh{\bm{\rho}}= \bm{\Phi}\bm{a}_0 \\ ~\\
\bm{P}_{\rho}= \sigma_{a}^2\bm{\Phi} \bm{\Phi}^t
\earr\right.
\\ ~\\
\sigma_{a}\longrightarrow \infty &&
\left\{\barr{@{}l@{}}
\wh{\bm{\rho}}= \bm{\Phi} [\bm{A}^t\bm{A}]^{-1}\bm{A}^t \bm{F} \\ ~\\
\bm{P}_{\rho}=\sigma_{\epsilon}^2 \bm{\Phi}[\bm{A}^t\bm{A}]^{-1}\bm{\Phi}^t
\earr\right.
\earr
\\ & ~\\ \hline
\earr
\]
Table 1: A comparison between parametric and non-parametric approaches.
}

When the posterior covariance matrix $\bm{P}_{\rho}$ is calculated, we can
use it to give some information about the uncertainty of the solution.
For example, we can use its diagonal
elements to calculate $\sigma_k=\sqrt{P_{kk}}$ and use it to error bars on
the solution.

\section{Numerical experiments}

In order to demonstrate the preceding
considerations  we make use of the following analytical
model. For a charge density given by a symmetric Fermi
distribution~\cite{GLMP91}
\beq\label{Trho}
\rho(r) =\alpha\,\frac{\cosh(R/d)}{\cosh(R/d)+\cosh(r/d)}
\eeq
an analytical expression for the corresponding charge form factor
can easily be obtained~\cite{K91,ACP94}:
\beq\label{TFc}
F(q) = -\frac{4\pi^2\alpha d}{q} \, \frac{\cosh(R/d)}{\sinh(R/d)}
\, \left[\frac{R\,\cos(qR)}{\sinh(\pi qd)} -\frac{\pi d\sin(qR)
\cosh(\pi qd)}{\sinh^2(\pi qd)}\right].
\eeq
Only two of the parameters $\alpha$, $R$ and $d$ are independent
since the charge density must fulfill the normalization condition
\beq\label{Norm}
4\pi\int r^2\,\rho(r)\d{r} = Z.
\eeq

Figure~1 shows the theoretical charge density $\rho(r)$
of ~$^{12}$C ~(Z=6) obtained from (\ref{Trho})
for $r\in[0,0.8]$ with $R=1.1$ A$^\frac{1}{3}$ and $d=0.626$ fm
and the theoretical charge form factor $F_{c}(q)$ obtained by
(\ref{TFc}) for $q\in[0,8]$ fm$^{-1}$ and the nine simulated
experimental data:
$$
\bm{q}=[0.001, .5, 1.0, 2.0, 3.0, 4.0, 5.0, 6.0, 7.0] \, \mbox{fm}^{-1}
$$
which are used as inputs to all the inversion methods.

\vspace{-10mm}
$$
\includegraphics[width=64mm,height=64mm]{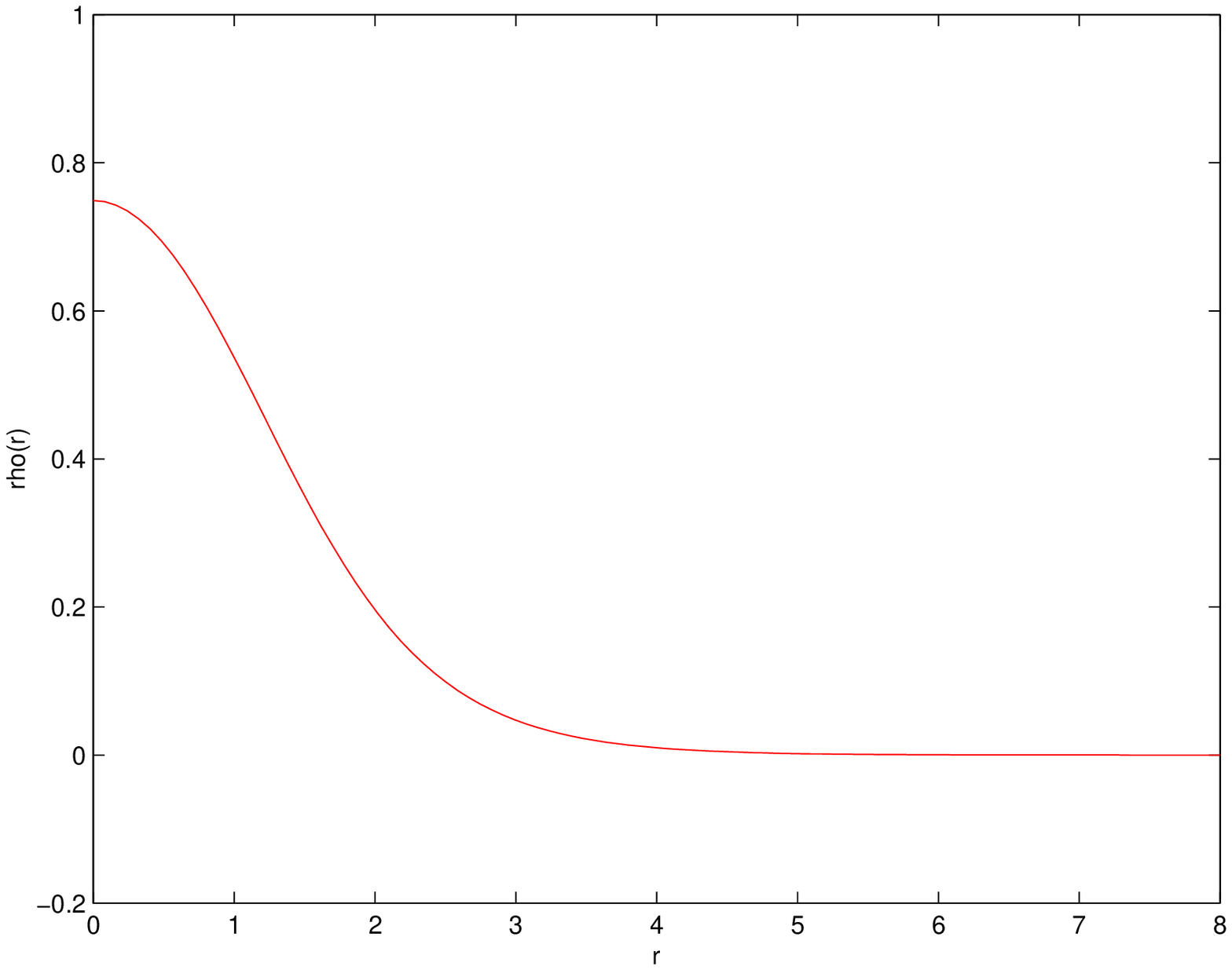}
\includegraphics[width=64mm,height=64mm]{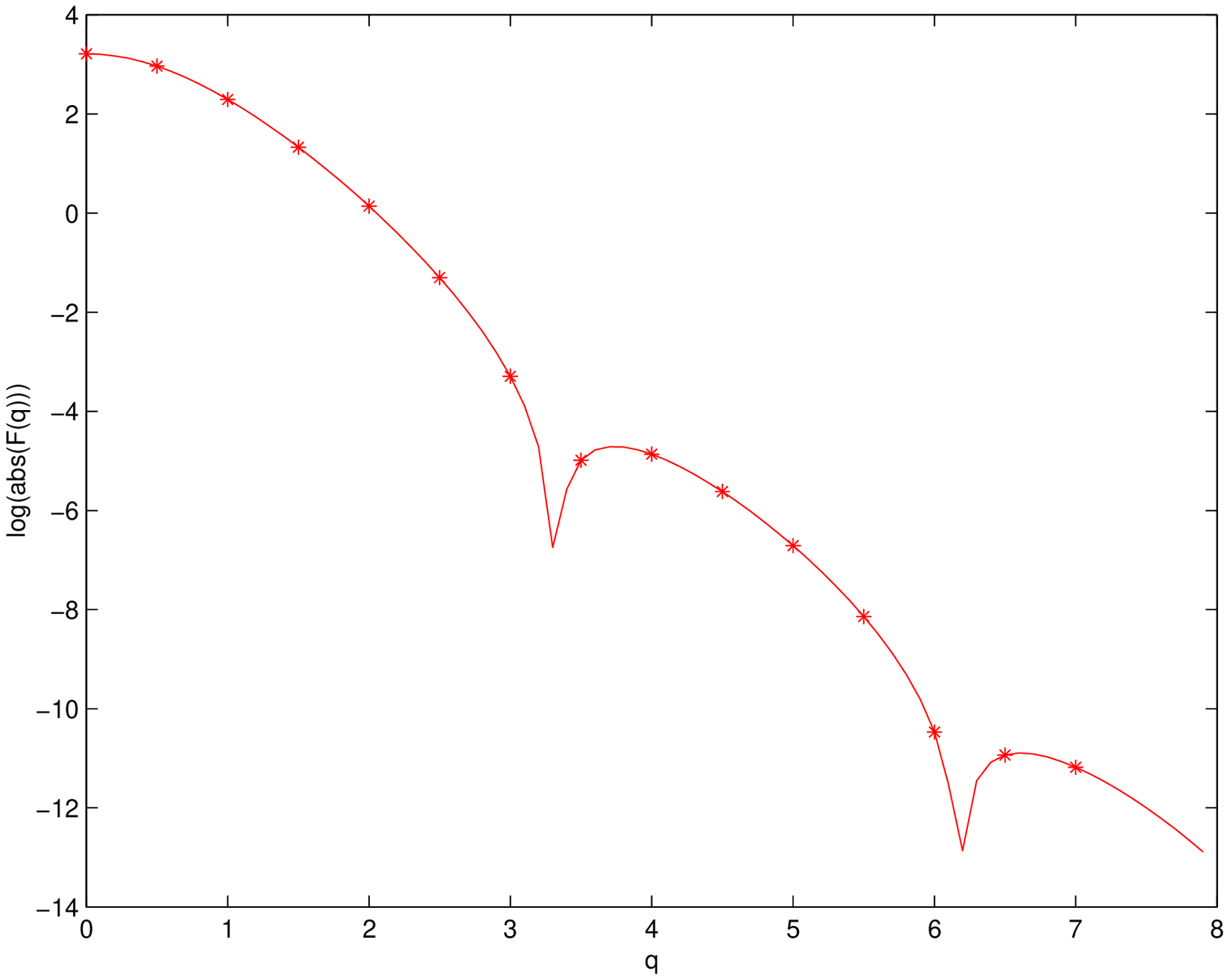}
$$
\noindent{\footnotesize
Figure~1. Theoretical charge density $\rho(r)$ [left],
charge form factor $\log\|F_{c}(q)\|$ and the data
[stars] used for numerical experiments [right].
}

\newpage
\subsection{Experiments with operator based parametric models}

We use these data in the parametric model (\ref{Exp1})
with $R_{c}=8$~fm and estimate the coefficients $\bm{a}$ by
\beqnn
\mbox{LS :} & &
\wh{\bm{a}}=(\bm{A}^t\bm{A})^{-1}\bm{A}^t\bm{F}_{c} \\
\mbox{MNLS :} & &
\wh{\bm{a}}
=(\bm{A}^t\bm{A}+\lambda\bm{I})^{-1}\bm{A}^t\bm{F}_{c} \\
\mbox{MAP :} & &
\wh{\bm{a}}
=(\bm{A}^t\bm{A}+\lambda\bm{D}^t\bm{D})^{-1}\bm{A}^t\bm{F}_{c}
\eeqnn
Then using these coefficients we calculate
$\rho(r)$ by (\ref{Exp1}) and $F_{c}(q)$ by (\ref{Fcexp1}).

\medskip\noindent
Figure~2 and Figure~3 show the reconstructed charge densities
$\wh{\bm{\rho}}$ and the corresponding charge form factors
$\wh{\bm{F}}_{c}$ obtained by LS and by MNLS
for $N=5$ and $N=10$.

\medskip\noindent
Figure~4 shows the reconstructed charge densities
by LS and by MNLS for different expansion order $N$ from $5$ to $10$.
Note that the LS solutions are very sensitive and
vary greatly with $N$, but the MNLS solution
stays more stable with respect to $N$.

\medskip\noindent
Figure~5 shows the reconstructed charge densities and the
corresponding charge form factors obtained by MNLS and MAP
for $N=30$.

$$
\includegraphics[width=64mm,height=64mm]{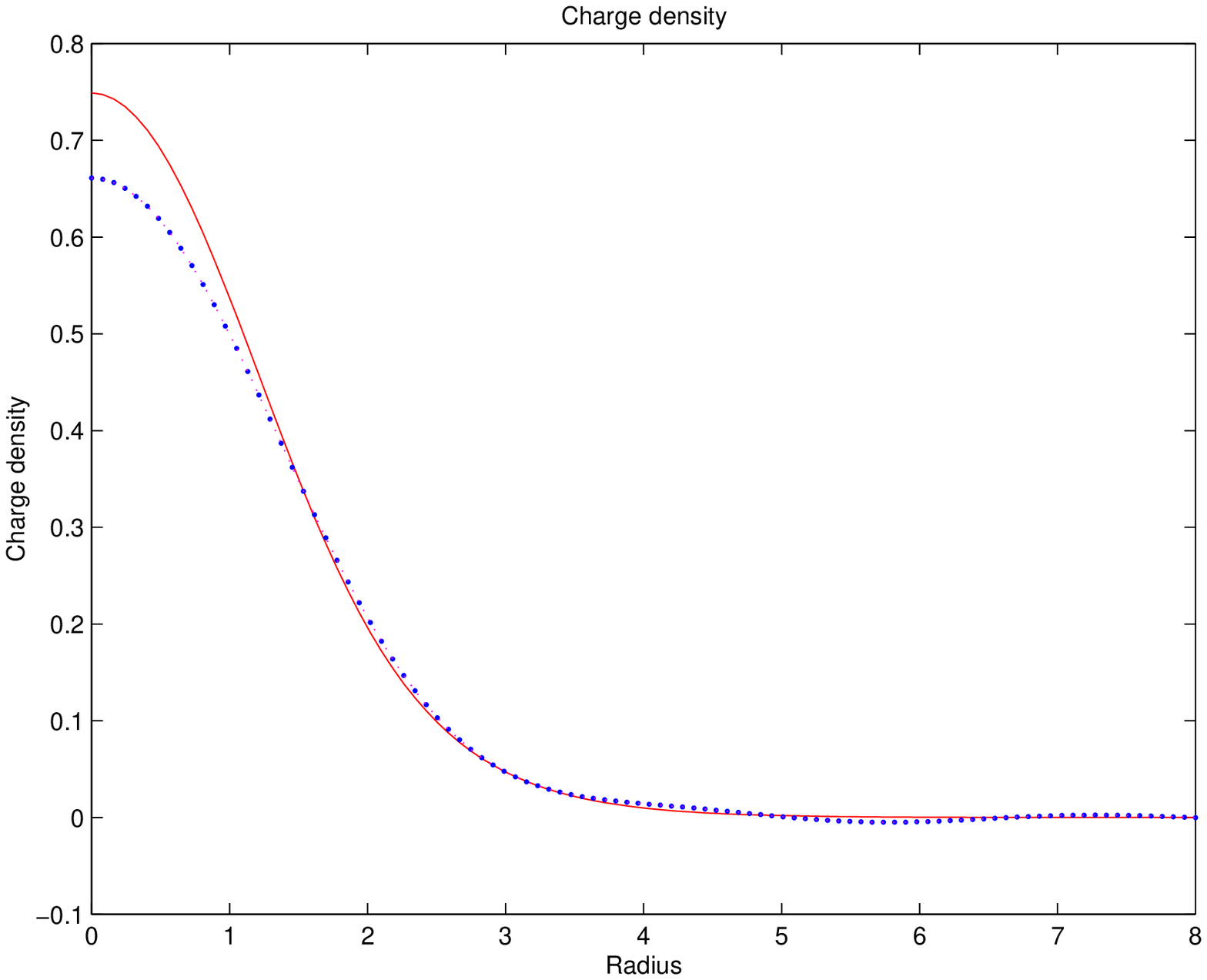}
\includegraphics[width=64mm,height=64mm]{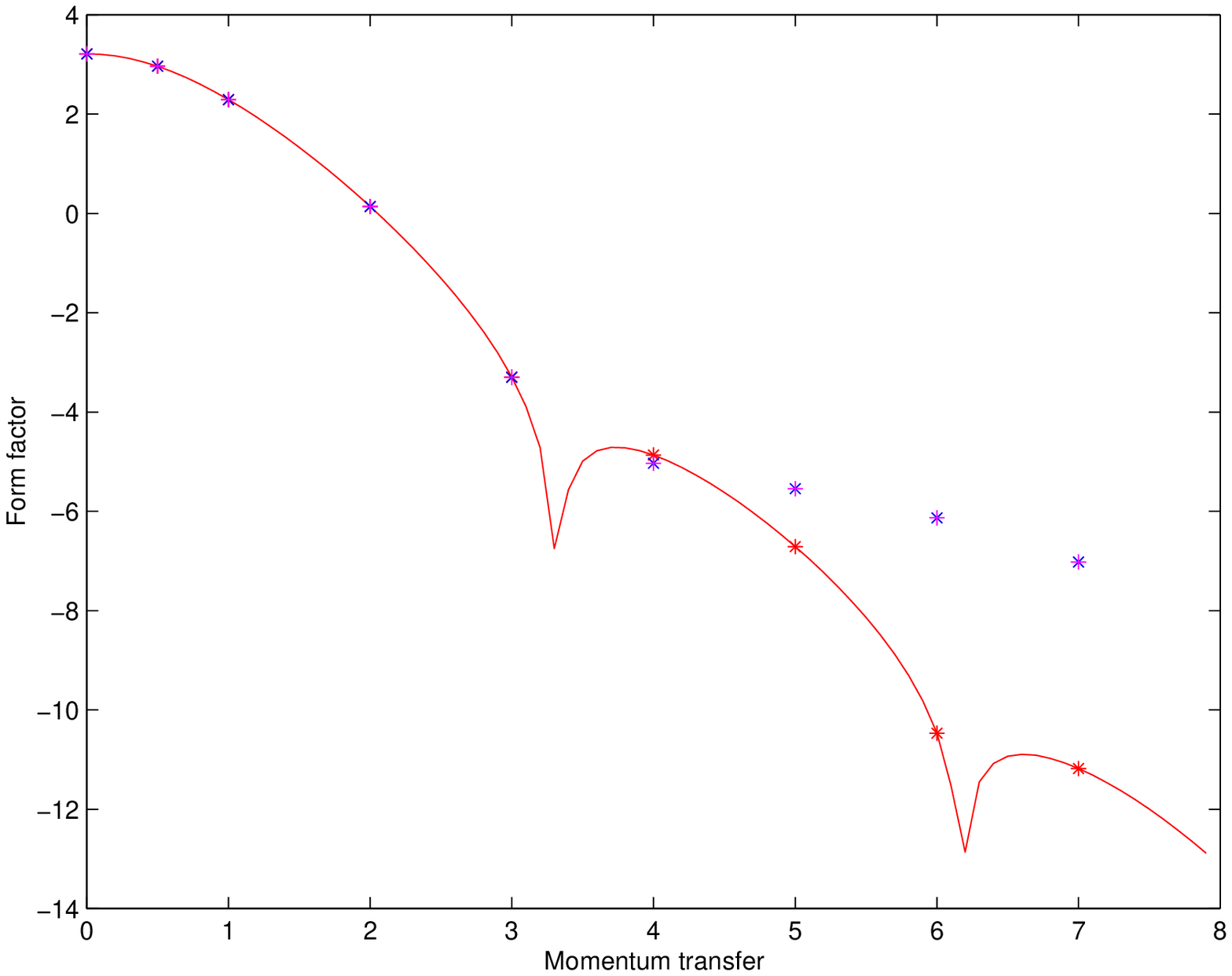}
$$
\noindent{\footnotesize
Figure~2. Parametric reconstruction of $\wh{\rho}(r)$
obtained by LS (point) and by MNLS (dotted) for $N=5$ [left]
and the corresponding reconstructed charge form factors
$\log\|\wh{F}_{c}(q)\|$ [right].
Two solutions are practically indistinguishable and both not very
satisfactory due to a large bias of the solution for small radius $r$.
Note also that both solutions fit well the data.
}

$$
\includegraphics[width=64mm,height=64mm]{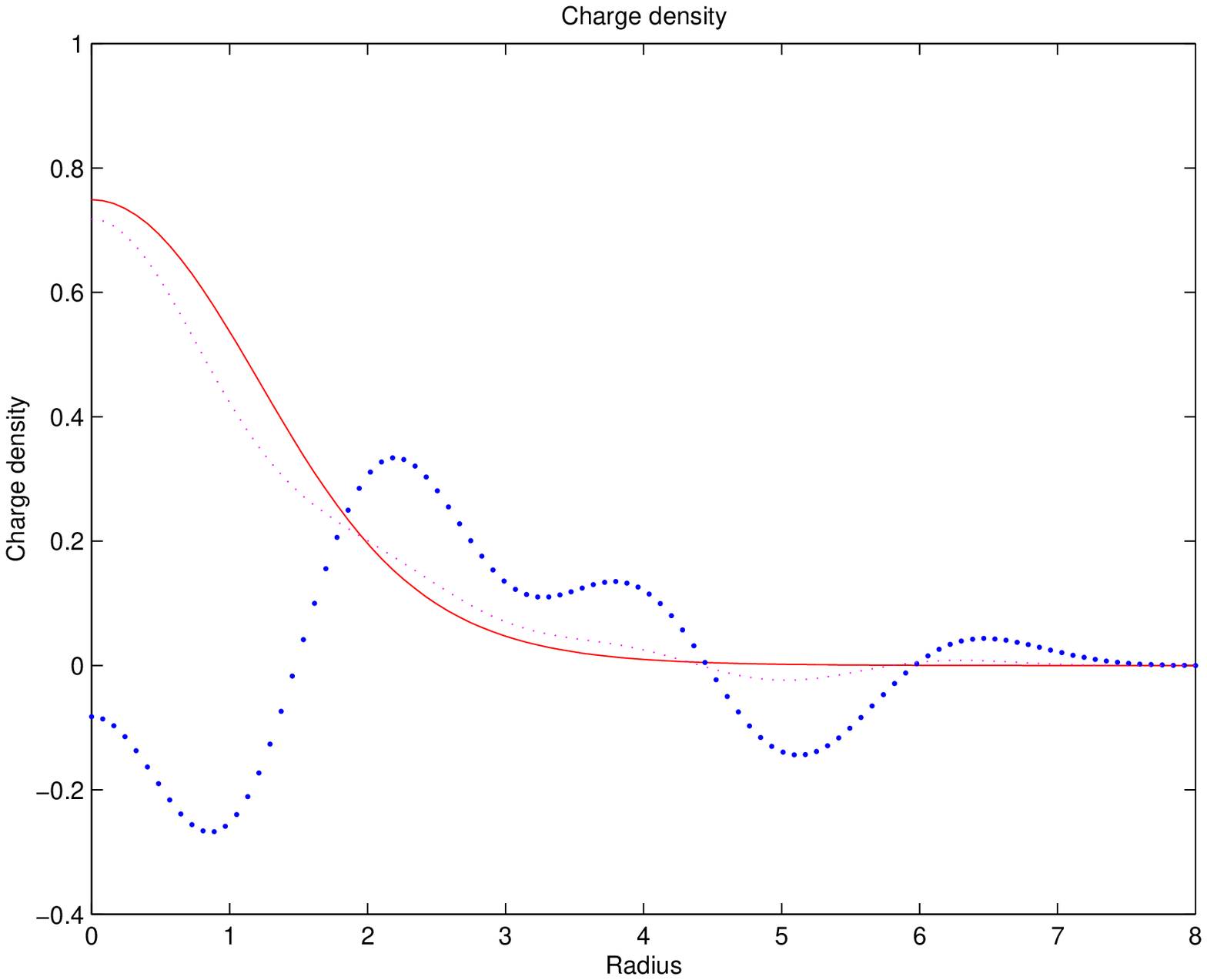}
\includegraphics[width=64mm,height=64mm]{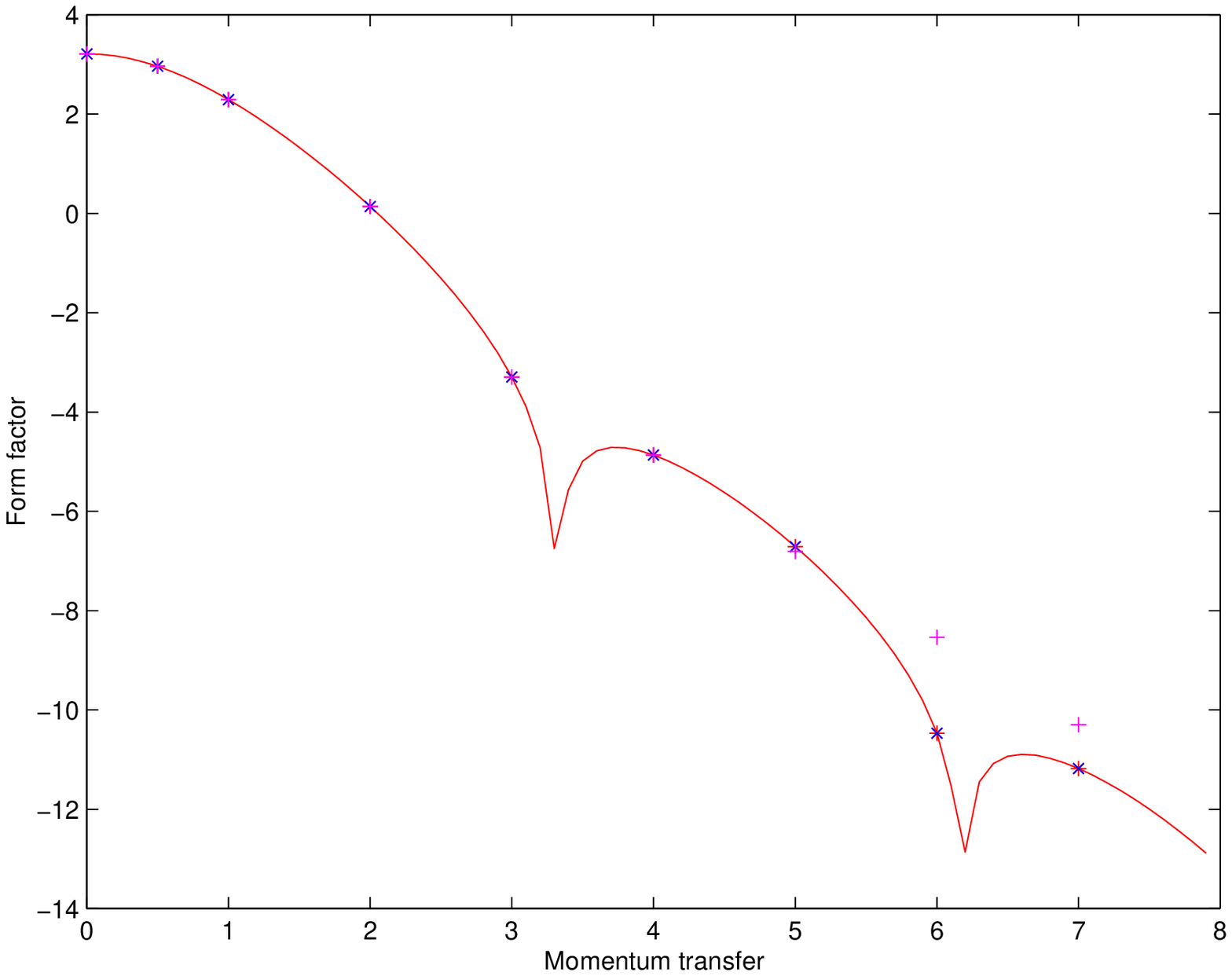}
$$
\noindent{\footnotesize
Figure~3. Parametric reconstruction of $\wh{\rho}(r)$
obtained by LS (point) and by MNLS (dotted) for $N=10$ [left]
and the corresponding reconstructed charge form factors
$\log\|{F}_{c}(q)\|$ [right].
Note that the LS solution fits very well the data but is very
unstable but the MNLS solution, which does not fit perfectly the data,
is at least more stable.
}

$$
\includegraphics[width=64mm,height=64mm]{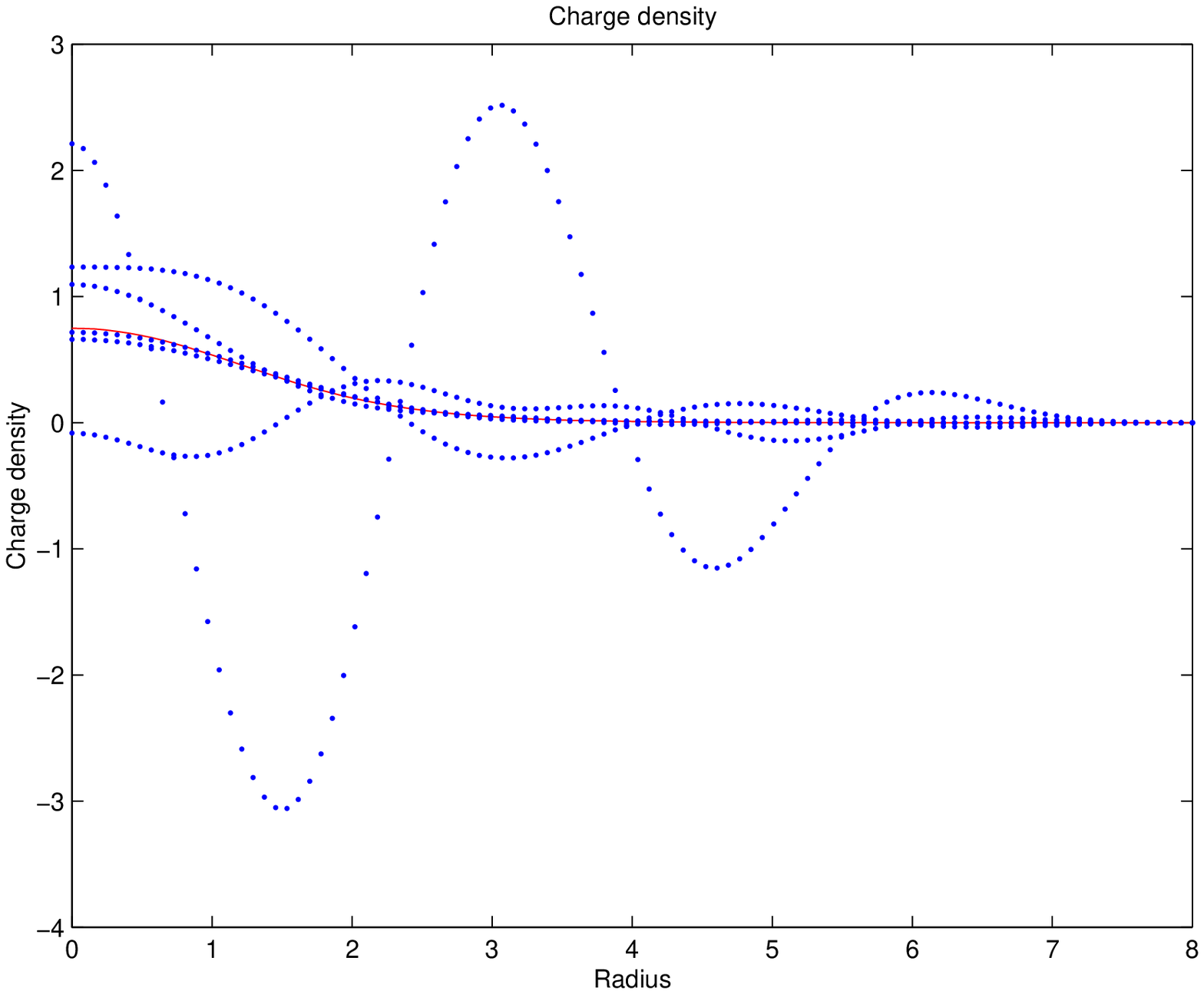}
\includegraphics[width=64mm,height=64mm]{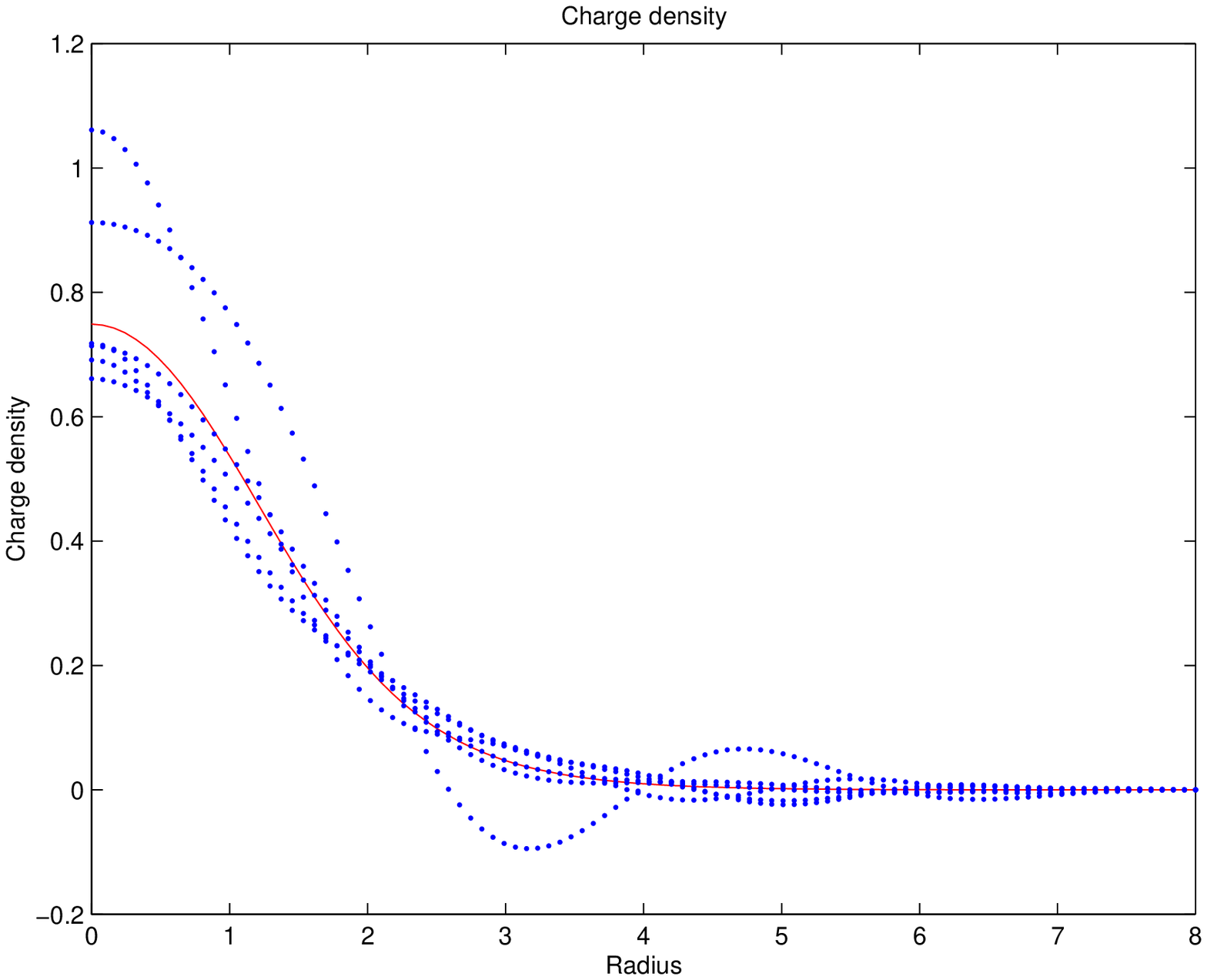}
$$
\noindent{\footnotesize
Figure~4. Parametric reconstruction of $\wh{\rho}(r)$
obtained by LS (left) and by MNLS (right) for different values of $N: =5:1:10$.
Note that the LS solutions vary greatly with $N$, but the MNLS solutions
stay more stable with respect to $N$.
}

$$
\includegraphics[width=64mm,height=64mm]{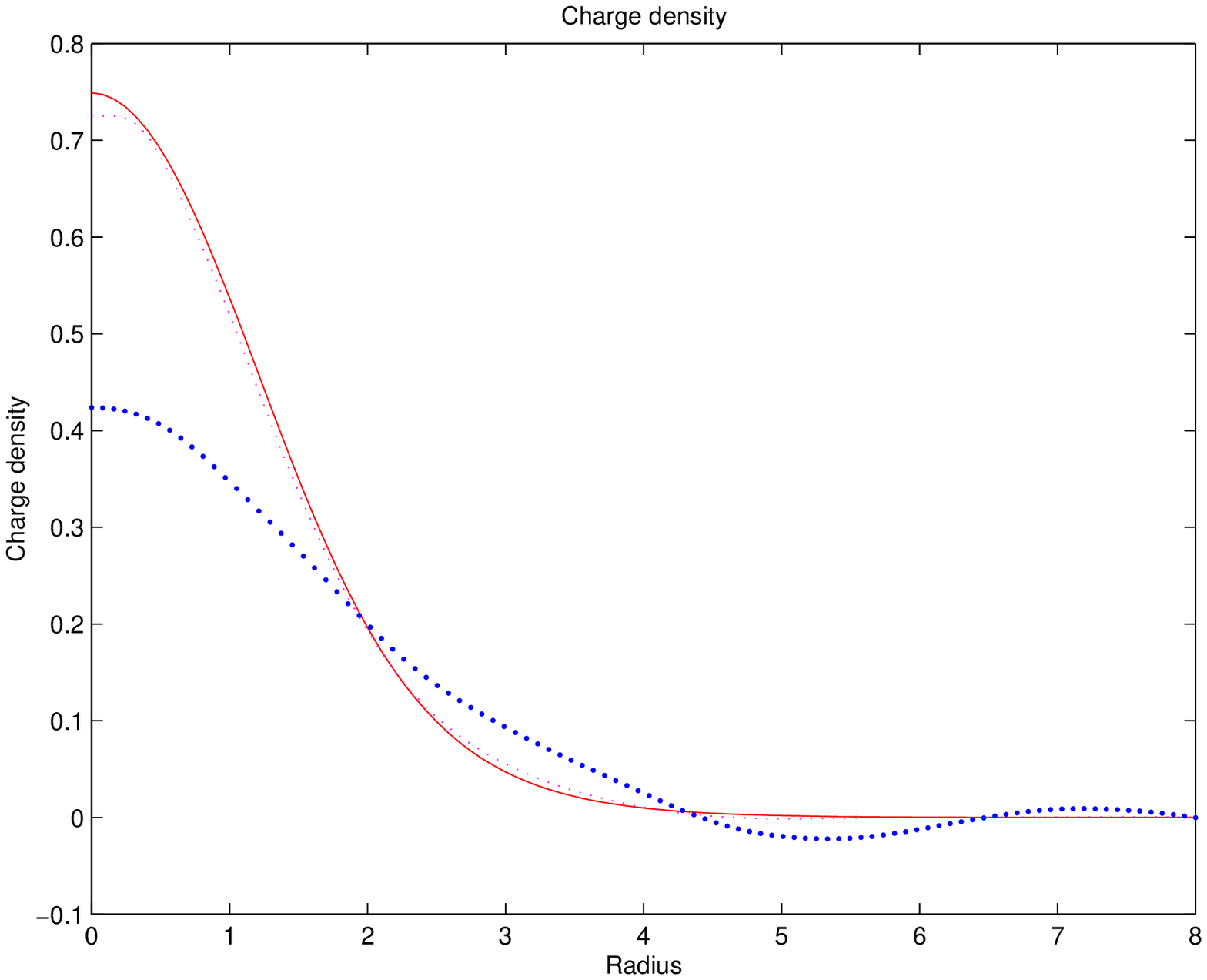}
\includegraphics[width=64mm,height=64mm]{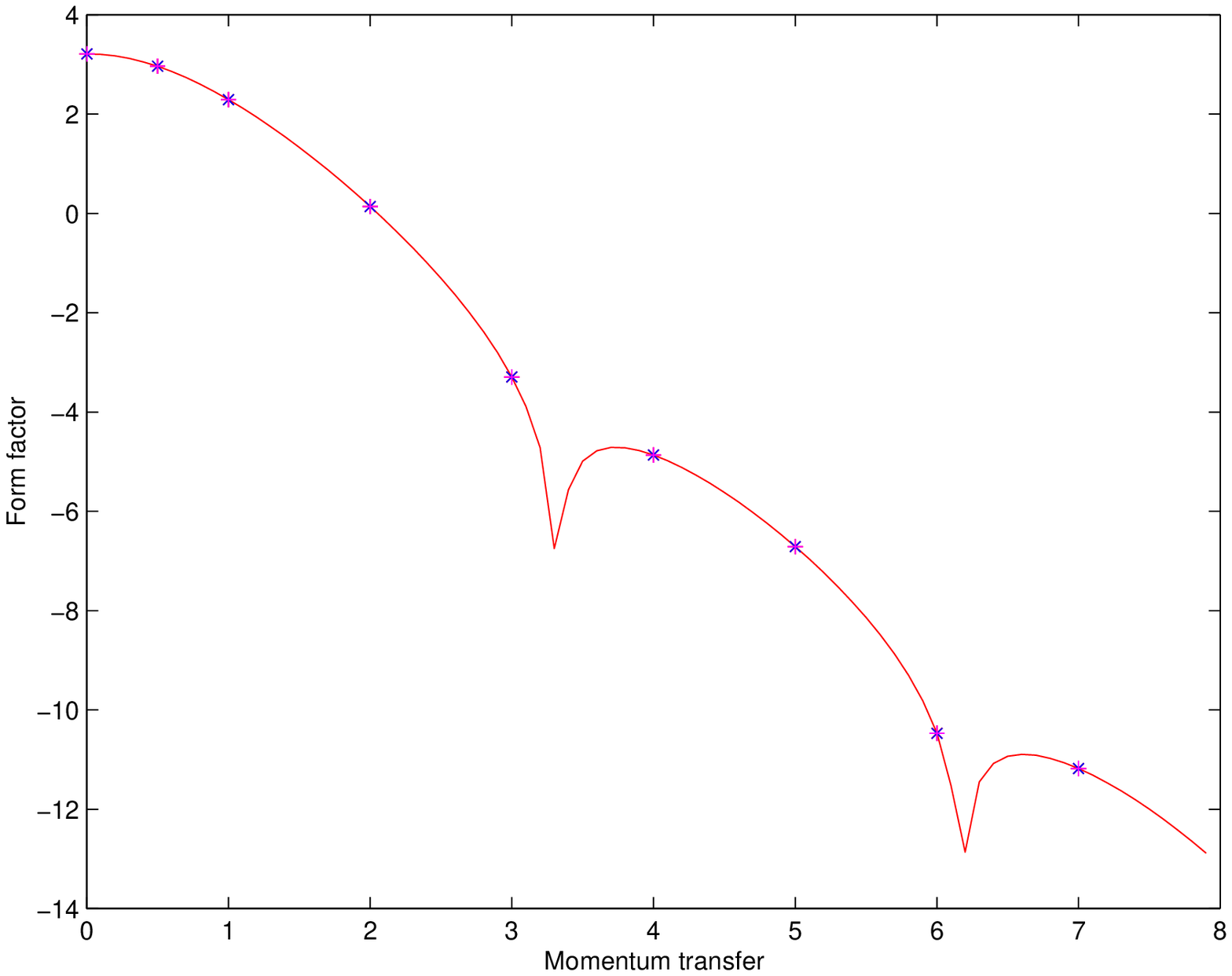}
$$
\noindent{\footnotesize
Figure~5. Parametric reconstruction of $\wh{\rho}(r)$
obtained by MNLS (point) and by MAP (dotted) for $N=30$
[left] and the corresponding reconstructed charge form factors
$\log\|\wh{F}_{c}(q)\|$ [right].
In this case the LS solution is completely unrealistic and is not
presented. The MNLS solution has a large bias for small raduis.
The MAP solution is very satisfactory. Note also that both solutions
satisfy the data constraint practically in the same way.
}

\vspace{2cm}
\subsection{Experiments with non parametric models}

The same data are then used with the non parametric model
(\ref{Discrete2}) with $N=100$ and $R_{c}=8$~fm and
$\wh{\bm{\rho}}$ is calculated by
\beqnn
\mbox{LS :} & &
\wh{\bm{\rho}}=(\bm{B}^t\bm{B})^{-1}\bm{B}^t\bm{F}_{c} \\
\mbox{MNLS :} & &
\wh{\bm{\rho}}
=(\bm{B}^t\bm{B}+\lambda\bm{I})^{-1}\bm{B}^t\bm{F}_{c} \\
\mbox{MAP1 :} & &
\wh{\bm{\rho}}
=(\bm{B}^t\bm{B}+\lambda\bm{D}^t\bm{D})^{-1}\bm{B}^t\bm{F}_{c} \\
\eeqnn

Figure~6 shows the estimated $\wh{\bm{\rho}}$ and the
corresponding $\wh{\bm{F}}_{c}$ by MNLS and by MAP.

Figure~7 shows two solutions obtained by a parametric and a
non-parametric
method and their associated error bars.

$$
\includegraphics[width=64mm,height=64mm]{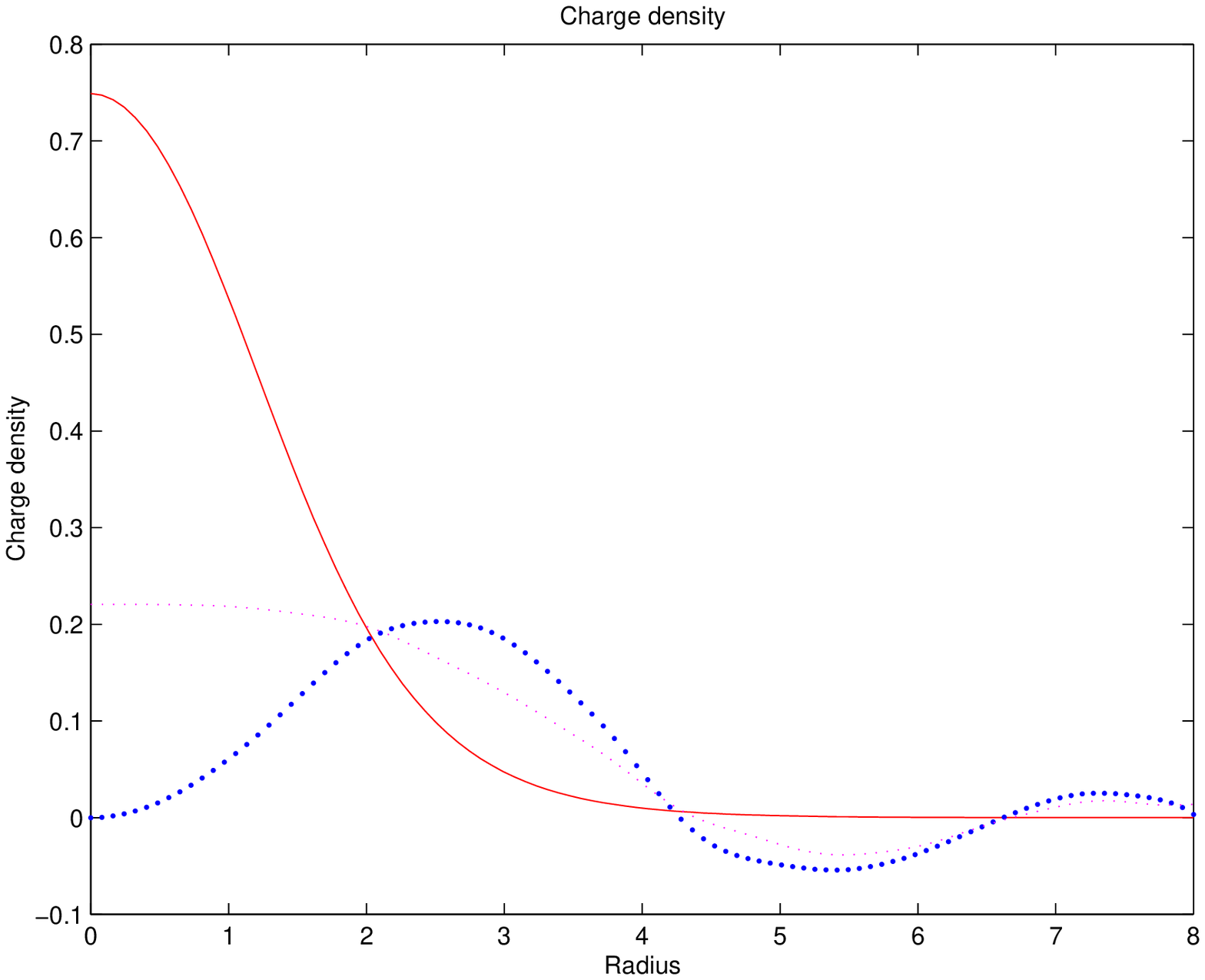}
\includegraphics[width=64mm,height=64mm]{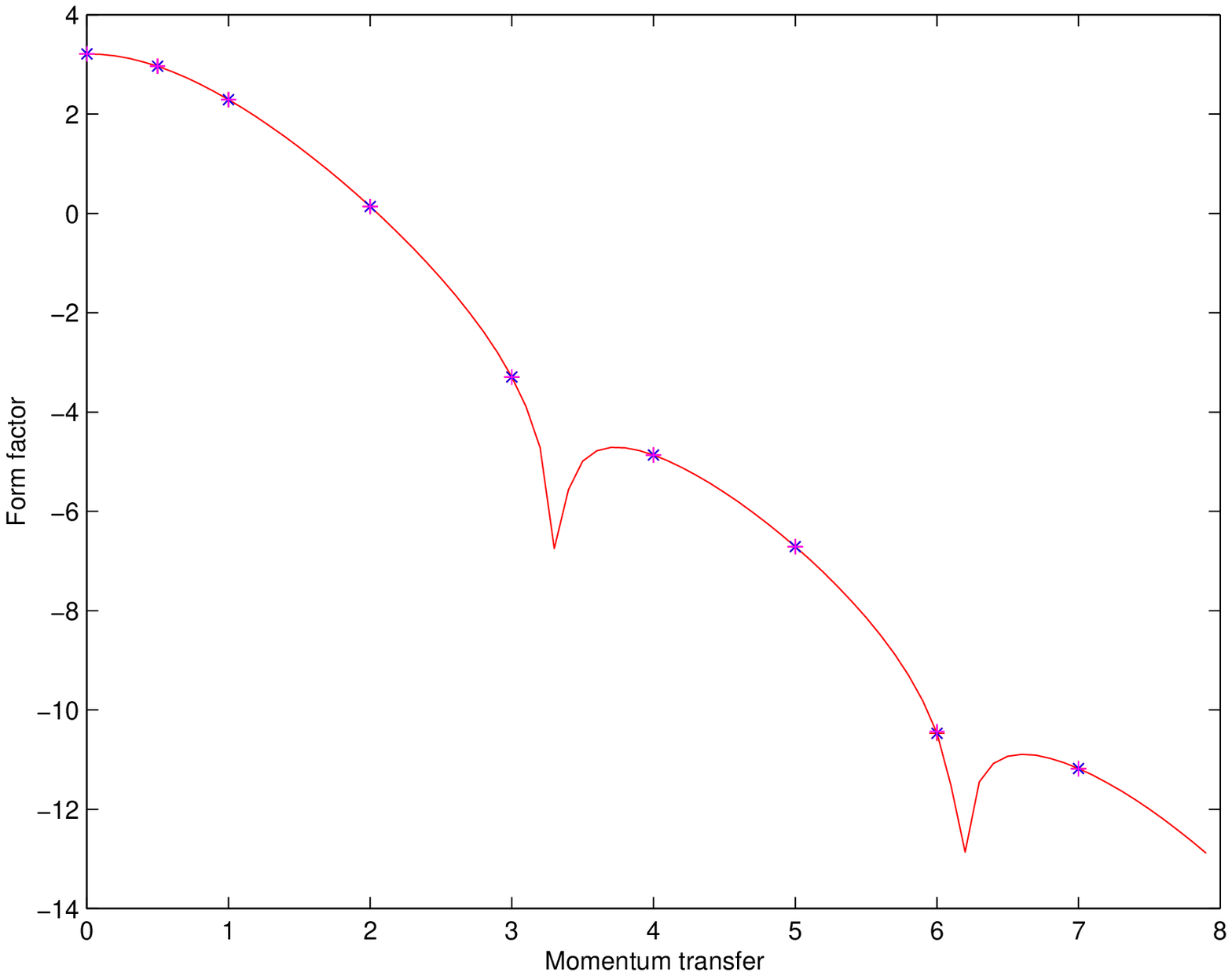}
$$
\noindent{\footnotesize
Figure~6. Non parametric reconstruction of
$\wh{\rho}(r)$ obtained by MNLS (point) and MAP1 (dotted)
for $N=100$ [left] and the corresponding reconstructed
charge form factors $\log\|\wh{F}_{c}(q)\|$ [right].
}

$$
\includegraphics[width=64mm,height=64mm]{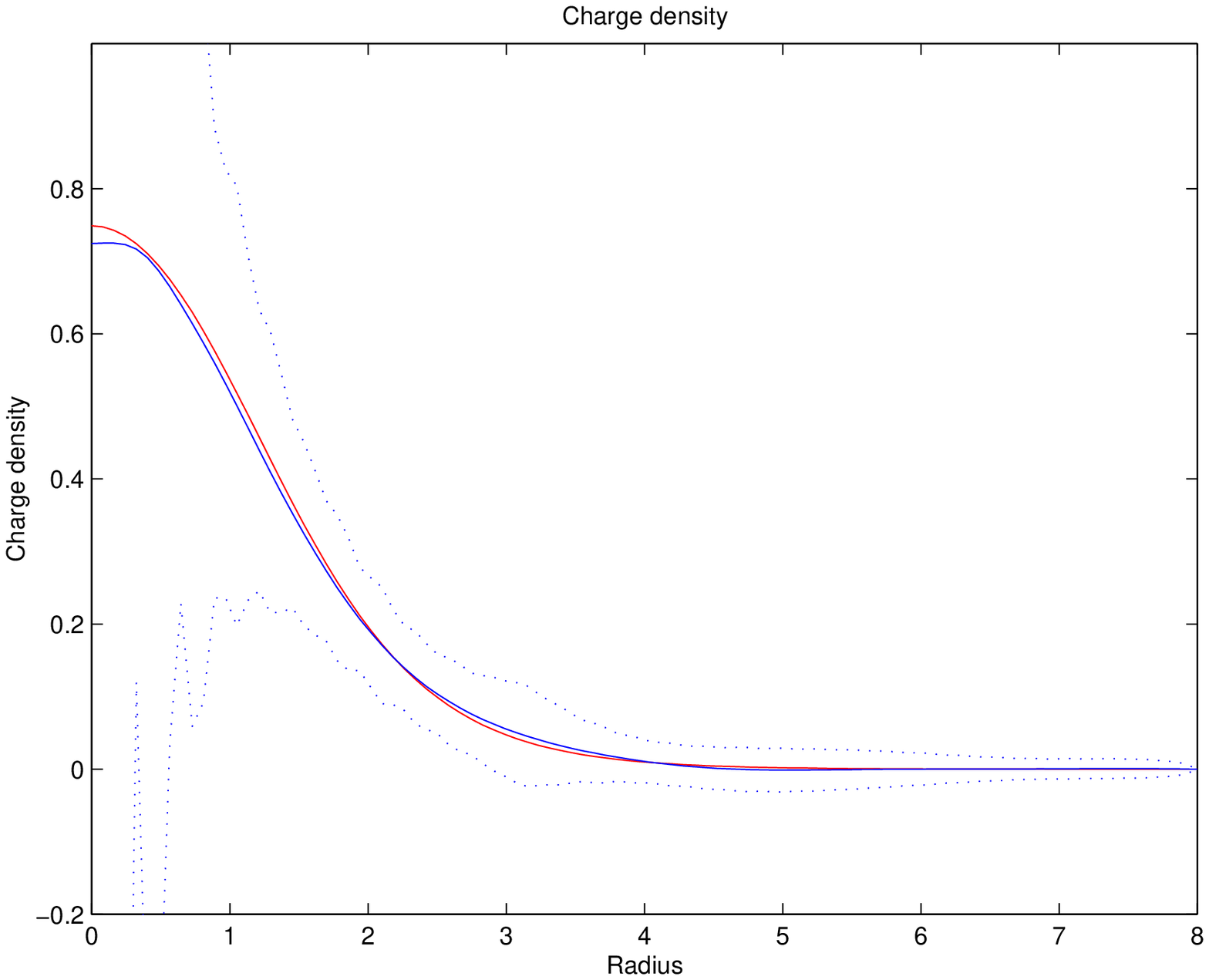}
\includegraphics[width=64mm,height=64mm]{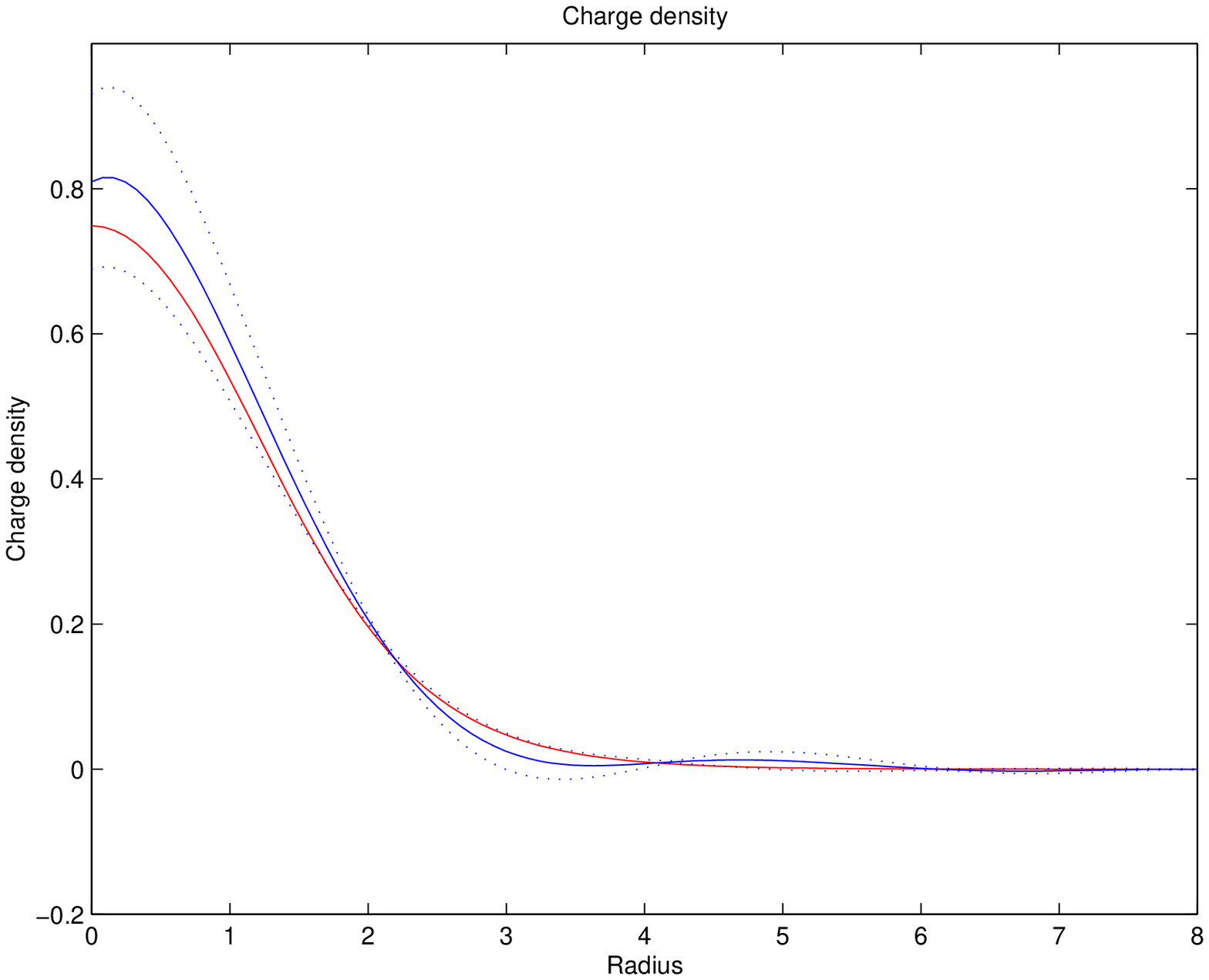}
$$
\noindent{\footnotesize
Figure~8. Uncertainty in parametric and non parametric methods:\\
Left: Parametric reconstruction of $\wh{\rho}(r)$
obtained by MAP1 \\
Right: Non-parametric reconstruction of $\wh{\rho}(r)$
obtained by MAP1
}

\newpage
\section{Conclusion}

We considered the problem of the determination of the charge density
from a limited number of charge form factor measures as
an ill-posed inverse problem.
We proposed a Bayesian probabilistic approach to this
problem and showed how many classical methods
can be considered as special cases of the proposed approach.
We addressed also the problem of the basis function
choice for the discretization and the uncertainty of the solution.
We illustrated the performances of the proposed methods by some numerical results.

\end{document}